\pdfoutput=1
\documentclass[reprint,
superscriptaddress,
 amsmath,amssymb,
 aps,
 nofootinbib,
]{revtex4-1}
\usepackage{color}
\usepackage{feynmf}
\usepackage{graphicx}
\usepackage{dcolumn}
\usepackage{bm}
\usepackage[utf8]{inputenc}
\usepackage{hyperref}
\usepackage{lipsum}
\usepackage{slashed}
\usepackage{float}
\usepackage{subcaption} 

\begin{document}


\title{Competition of inhomogeneous chiral phases and two-flavor color superconductivity in the NJL model}

\author{Phillip Lakaschus}
\affiliation{Institute for Theoretical Physics, Goethe University, Max-von-Laue-Str.\ 1,
D-60438 Frankfurt am Main, Germany}
\author{Michael Buballa}
\affiliation{Technische Universit\"at Darmstadt, Department of Physics, Institut f\"ur Kernphysik, Theoriezentrum, Schlossgartenstr.\ 2, D-64289 Darmstadt, Germany}
\affiliation{Helmholtz Research Academy Hesse for FAIR, Campus Riedberg, Max-von-Laue-Str.\ 12, 
D-60438 Frankfurt am Main, Germany}

\author{Dirk H.\ Rischke}
\affiliation{Institute for Theoretical Physics, Goethe University, Max-von-Laue-Str.\ 1,
D-60438 Frankfurt am Main, Germany}
\affiliation{Helmholtz Research Academy Hesse for FAIR, Campus Riedberg, Max-von-Laue-Str.\ 12, 
D-60438 Frankfurt am Main, Germany}

\date{\today}

\begin{abstract}
We study the phase structure of the 
two-flavor Nambu--Jona-Lasinio (NJL) model
\textcolor{black}{in the chiral limit},  extending a previous study of the competition of an inhomogeneous chiral phase and a two-flavor color-superconducting (2SC) phase \cite{Sadzikowski:2002iy,Sadzikowski:2006jq}. There,
an analytic expression for the dispersion relations for quasiparticle excitations in the presence of both a particular inhomogeneous chiral condensate, the so-called chiral density wave (CDW), and a homogeneous 2SC condensate was found. In this work we show how to determine the dispersion relations for arbitrary modulations of the chiral condensate in the presence of a homogeneous 2SC condensate, if the dispersion relations in the absence of color superconductivity are known. 
In our calculations, we employ two different Ans\"atze for the inhomogeneous chiral condensate, the CDW as well as the real-kink crystal (RKC). 
Depending on the value of the diquark coupling we find a region of the phase diagram where the inhomogeneous chiral and the 2SC condensates coexist, 
confirming results of Refs.\  \cite{Sadzikowski:2002iy,Sadzikowski:2006jq}. Decreasing the diquark coupling favors the inhomogeneous phase over the coexistence phase. On the other hand, increasing the diquark coupling leads to a larger 2SC phase, while the inhomogeneous chiral and the coexistence phases become smaller.
In agreement with previous studies the RKC Ansatz is  energetically preferred over the CDW Ansatz. Both Ans\"atze lead to a qualitatively similar phase diagram, however the coexistence phase is smaller for the RKC Ansatz.
\end{abstract}

\pacs{Valid PACS appear here}
\maketitle


\section{Introduction}
\label{sec:level1}

The exploration of the QCD phase diagram is one of the major topics in contemporary high-energy nuclear physics. The chiral symmetry of QCD, which is broken in the vacuum and restored at high temperatures and densities, plays an important role in determining the structure of the QCD phase diagram.
Lattice-QCD studies have shown that chiral symmetry is restored in a crossover transition at a temperature $T \sim (156.5 \pm 1.5)$ MeV at physical values of the quark mass \cite{Bazavov:2018mes}. However, due to the fermion sign problem these methods are not applicable for nonvanishing quark chemical 
potentials $\mu$. Several calculations within continuum approaches suggest that the 
transition becomes of first order
at large $\mu$ and moderate $T$, which terminates at a second-order critical endpoint [for a compilation of recent results for the location of this endpoint, see
Ref.\ \cite{Gao:2020fbl}]. 

Calculations in various QCD-inspired models, like the Gross-Neveu (GN)
\cite{Thies:2006ti}, the
Nambu--Jona-Lasinio (NJL) \cite{Sadzikowski:2000ap,Nakano:2004cd, Nickel:2009wj}, the quark-meson (QM)
\cite{Nickel:2009wj,Carignano:2014jla,Adhikari:2017ydi}, and the parity-doublet model \cite{Heinz:2013hza}, suggest that, within mean-field approximation, the first-order chiral phase transition is superseded by an inhomogeneous phase, 
where the chiral condensate varies as a function of 
spatial 
coordinates, for a review, see Ref.\ \cite{Buballa:2014tba}. In 
\textcolor{black}{most of} these studies, specific Ans\"atze for the shape of the chiral order parameter are chosen that allow for an analytic treatment of the problem, such as the chiral density wave (CDW) or the real-kink crystal (RKC) Ansatz.
Alternative approaches to inhomogeneous chiral phases,
which are not based on a specific Ansatz for the shape of the condensate, \textcolor{black}{are,} e.g., Ginzburg--Landau theory
\cite{Abuki:2011pf,Nickel:2009ke} and studies where the shape of the condensate is determined by minimization of the action of the theory in mean-field approximation \cite{Heinz:2015lua}.
The existence of an inhomogeneous phase was recently confirmed in calculations
within the 1+1--dimensional GN model in full quantum field theory on the lattice \cite{Lenz:2020bxk}. 
At asymptotically large $\mu$ and moderate $T$, asymptotic freedom predicts that quark matter is a color superconductor \cite{Alford:1997zt,Rischke:2003mt,Alford:2007xm}. It is an interesting open question, how far this color-superconducting phase extends towards lower $\mu$ and whether it competes with the inhomogeneous chiral phase.

The only works which, to our knowledge, have addressed this question are those of Refs.\ \cite{Sadzikowski:2002iy,Sadzikowski:2006jq,Nowakowski:2016dwu}. References \cite{Sadzikowski:2002iy,Sadzikowski:2006jq} found
a solution for the dispersion relations of quasiparticle excitations in the presence of both
a CDW-modulated chiral condensate and a homogeneous 2SC condensate. A region in the phase diagram was identified where an inhomogeneous chiral condensate and a 2SC diquark condensate coexist\footnote{Phase coexistence between a
homogeneous chiral and a 2SC
color-superconducting phase has been found in Ref.\ \cite{Blaschke:2003cv}.}. However, the Lifshitz and \textcolor{black}{the} tricritical 
point 
do not coincide, which contradicts the Ginzburg--Landau studies of
Ref.\ \cite{Nickel:2009ke}. A possible source for this discrepancy could be the particular regularization scheme used in Refs.\ \cite{Sadzikowski:2002iy,Sadzikowski:2006jq}, which is based on a Taylor expansion of the grand potential in the wave number $q$.
Reference \cite{Nowakowski:2016dwu} confirmed the existence of a coexistence phase at $T=0$ using the Pauli-Villars regularization scheme and extended this study to nonvanishing isospin chemical potential.

In this work, we extend Refs.\ \cite{Sadzikowski:2002iy,Sadzikowski:2006jq,Nowakowski:2016dwu} in the following ways. First, we investigate the phase diagram at finite $T$ and $\mu$ (but at
vanishing isospin chemical potential) using
the Pauli-Villars regularization. Second, we derive a method to compute the quasi-particle dispersion relations for arbitrary modulations of the chiral condensate in the presence of a homogeneous 2SC condensate, if the dispersion relations in the absence of the latter are known. We subsequently apply this method to the RKC in addition to the CDW. 

This paper is organized as follows: In Sec.\ 
\ref{sec:model} we present the NJL model and extend it by a quark-quark interaction term that corresponds to the spin-zero color-antitriplet channel and allows for 2SC color-superconducting condensates. We then present the method to compute the quasi-particle dispersion relations for arbitrary modulations of the chiral condensate in the presence of a 2SC color-superconducting condensate and apply this to compute the grand potential for the CDW and the RKC Ansatz, respectively. 
In Sec.\ \ref{sec:results} we present the phase diagrams for the two inhomogeneous Ans\"atze and different values of the diquark coupling. We conclude this work with a summary and an outlook in Sec.\ \ref{sec:discussion}.

\section{The Model}
\label{sec:model}

\subsection{The NJL model with diquarks}
\label{sec:njl}

We consider the Lagrangian
\begin{equation} \label{eq:Lagr}
 \mathcal{L}_{\textrm{NJL+}\Delta} = \mathcal{L}_\textrm{NJL} + \mathcal{L}_\Delta\,,   
\end{equation}
where
\begin{equation} \label{eq:Lagr_NJL}
    \mathcal{L}_{\textrm{NJL}} = \bar{\psi} (i \gamma^\mu \partial_\mu + \mu \gamma_0 ) \psi + G \left[ (\bar{\psi} \psi)^2 + (\bar{\psi} i \gamma_5 \vec{\tau} \psi)^2 \right] 
\end{equation}
is the standard Lagrangian of the NJL model for 
$N_{f}=2$ quark flavors \textcolor{black}{and $N_{c}=3$
color degrees of freedom}
in the chiral limit and at
finite quark chemical potential $\mu$, with
$\psi$ being a 4$N_{c}N_{f}\,$-dimensional quark spinor, $G$ the four-fermion coupling, $\gamma^{\mu}$ the Dirac matrices, and $\vec{\tau} = (\tau_1, \tau_2, \tau_3)$ the vector of Pauli matrices in flavor space.
The second term in Eq.\ (\ref{eq:Lagr}) is added in order to describe diquark condensation in the spin-zero color-antitriplet channel, 
\begin{equation}
    \mathcal{L}_{\Delta} = G_{\Delta} \left(\bar{\psi}_c i \gamma_5 \tau_2 \lambda_A \psi\right) \left(\bar{\psi} i \gamma_5 \tau_2 \lambda_A \psi_c\right) \, ,
\end{equation}
where $\psi_c = C \bar{\psi}^{\textrm{T}}$, with $C= i \gamma^2 \gamma^0$ being the charge-conjugation matrix, and $\lambda_{A} , \, A = 2, 5, 7$, are the antisymmetric Gell-Mann matrices in color space. 

We bosonize the action of the model by performing a Hubbard-Stratonovich transformation and double the fermion degrees of freedoms in such a way that it leaves the total path integral unchanged \cite{Schmitt:2014eka}. This yields the effective Lagrangian
\begin{align}
    \mathcal{L}_{\textrm{eff}} = \dfrac{1}{2}  & \Bigg[ \bar{\psi} (i  \slashed{\partial} + \mu \gamma_0 + \sigma 
    + i \gamma_5 \vec{\pi} \! \cdot \! \vec{\tau}) \psi \notag \\
    &+ \bar{\psi}_c (i \slashed{\partial} - \mu \gamma_0 + \sigma    
    + i \gamma_5 \vec{\pi} \! \cdot \! \vec{\tau}) \psi_c \notag \\
    & + \Delta_A \left(\bar{\psi}_c i \gamma_5 \tau^2 \lambda^A \psi\right) + \Delta_A^* \left(\bar{\psi} i \gamma_5 \tau^2 \lambda^A \psi_c\right)
    \notag \\ 
    & - \dfrac{\sigma^2 \! + \! \vec{\pi}^2}{2 G} - \dfrac{|\Delta_A|^2}{2 G_\Delta} \! \Bigg] \, \notag \\
    \equiv \dfrac{1}{2} & \left( \bar{\Psi} \mathcal{S}^{-1} \Psi - \dfrac{\sigma^2 \! + \! \vec{\pi}^2}{2 G} - \dfrac{|\Delta_A|^2}{2 G_\Delta} \right)\, ,
\end{align}
\textcolor{black}{with the real auxiliary fields
$\sigma$ and $\vec{\pi}$, corresponding to scalar- and pseudoscalar-meson degrees of freedom, and the complex auxiliary fields
$\Delta_A$, corresponding to diquarks.}
In the last step we introduced Nambu-Gor'kov spinors $\Psi = (\psi, \psi_c)^{\textrm{T}}$
and defined the inverse propagator
\begin{align}
    \mathcal{S}^{-1} = \begin{pmatrix} i \slashed{\partial}  - \hat M + \mu \gamma^0 & \hat\Delta \\ 
    -\hat \Delta^\dagger & i \slashed{\partial} -\hat M - \mu \gamma_0  \end{pmatrix}\, ,
\end{align}
with
\begin{equation}
      \hat M = - \sigma - i\gamma_5 \vec{\tau} \cdot \vec{\pi}
\end{equation}
and 
\begin{equation}
       \hat\Delta = i\gamma_5\tau_2\lambda_2 \Delta    
       \, .
\end{equation}

\subsection{Grand potential}
\label{sec:grandCanPot}

For spatially varying scalar and pseudoscalar fields 
$\sigma(\mathbf{x}),\, \vec{\pi}(\mathbf{x})$ and a constant color-superconducting gap parameter $\Delta_A = \delta_{A2} \Delta$, the grand potential reads in mean-field approximation
\begin{equation}\label{eq:Omega}
        \Omega = \Omega_\mathrm{kin}
        + \dfrac{1}{4 G} \frac{1}{V} \int d^3\mathbf{x} \big[ \sigma^2(\mathbf{x}) + \vec{\pi}^{\,2}(\mathbf{x}) \big] + \dfrac{|\Delta|^2}{4 G_\Delta}  \, ,
\end{equation}	
where 
\begin{equation} \label{eq:Omega_kin}
    \Omega_\mathrm{kin} \equiv -\frac{1}{2}\dfrac{T}{V} \textrm{Tr} \, \textrm{ln} \, \left( \dfrac{\mathcal{S}^{-1}}{T} \right) \,.
\end{equation}
Here, the functional trace is taken over space-time, spin, color, and flavor. The factor $1/2$ in front of the trace corrects for the artificial doubling of the quark degrees of freedom in the Nambu-Gor'kov formalism.

In order to evaluate the functional trace, we follow the approach of Refs.\ \cite{Nickel:2009wj,Carignano:2010ac} and isolate the time derivative
\begin{equation}
      \mathcal{S}^{-1} = \gamma^0 \big( i\partial_0 - H_\mathrm{NG} \big)\, ,
\end{equation}	
with the effective Dirac Hamiltonian
\begin{equation}
       H_\mathrm{NG} 
       =    
       \left( \begin{matrix} H - \mu & -\gamma^0 \hat\Delta \\ 
    \gamma^0 \hat \Delta^\dagger & H +\mu   \end{matrix} \right) \,,
\end{equation}	
where
\begin{equation}
      H \equiv -i \gamma^0\textrm{\boldmath $\gamma$} \cdot \nabla + \gamma^0 \hat M 
\end{equation}
is the effective Hamiltonian in the case without diquark pairing. Equation (\ref{eq:Omega_kin}) can then be written as
\begin{equation}\label{eq:Trln}
        \Omega_\mathrm{kin}        = 
         -\frac{T}{2 V} \sum_{n} \sum\limits_\lambda \mathrm{ln} \left( \dfrac{i \omega_n + \mathcal{E}_\lambda}{T} \right)  \, ,
\end{equation}
where $\mathcal{E}_\lambda$ are the eigenvalues of $H_\mathrm{NG}$.

The Matsubara sum can be evaluated with standard techniques \cite{kapusta} and yields
\begin{equation}\label{eq:cosh}
   \textcolor{black}{T}      \sum_{n}  \mathrm{ln} \left( \dfrac{i \omega_n + \mathcal{E}_\lambda}{T} \right)  = \dfrac{|\mathcal{E}_\lambda|}{2} + T \, \mathrm{ln} \left( 1 + \mathrm{e}^{-|\mathcal{E}_\lambda|/T} \right) \, .
\end{equation}
We determine the eigenvalues $\mathcal{E}_\lambda$ by squaring $H_{\textrm{NG}}$, which yields the block-diagonal matrix
\begin{equation}\label{eq:HNG2}
        H_\mathrm{NG}^2 
       = 
       \left( \begin{matrix} (H - \mu)^2  +   |\Delta|^2 \mathcal{P}_{rg}& 0  \\ 
       0  & (H +\mu)^2  + |\Delta|^2 \mathcal{P}_{rg} \end{matrix} \right) \,.
\end{equation}
Here
\begin{equation}
        \mathcal{P}_{rg} = \lambda_2^2 = \left( \begin{matrix} 1 & 0 & 0 \\ 0 & 1 & 0 \\ 0 & 0 & 0  \end{matrix} \right)
\end{equation}    
is the projector onto the space of gapped quark colors. 

The eigenvalues can now be read off from 
\textcolor{black}{those of}
the squared Hamiltonian, which are $(E_\lambda \mp \mu)^2$ for the ungapped quark/antiquark and, with twofold degeneracy,
 $(E_\lambda \mp \mu)^2 + |\Delta|^2$ for the gapped quarks/antiquarks, where $E_\lambda$ are the eigenvalues of $H$. The absolute values $\epsilon_{\lambda, \pm} \equiv |\mathcal{E}_{\lambda, \pm}|$ for quarks/antiquarks are thus found to be
 \begin{alignat}{2}
   \epsilon_{\lambda,\pm} &= \sqrt{(E_\lambda \mp \mu)^2 + |\Delta|^2}
   \quad & &\text{for the gapped quarks,}
   \label{eq:E_gapped}
\\   
    \epsilon_{\lambda,\pm}^{(0)}  &= |E_\lambda \mp \mu|
   \quad & &\text{for the ungapped quark.}  
   \label{eq:E_ungapped}
 \end{alignat}
Inserting these into Eq.\ (\ref{eq:Trln}) yields the expression
 \begin{align}
        \Omega_\mathrm{kin} 
        = 
         -\dfrac{1}{2} \sum\limits_{i=\pm}   \dfrac{1}{V} \sum\limits_\lambda 
         \Big[ 
         &\epsilon_{\lambda,i} + 2T \, \textrm{ln} \, \Big( 1 + \mathrm{e}^{-\epsilon_{\lambda,i}/T}\Big) \notag \\
         + &\frac{\epsilon^{(0)} _{\lambda,i}}{2} + T \, \textrm{ln} \, \Big( 1 + \mathrm{e}^{-\epsilon^{(0)} _{\lambda,i}/T}\Big)
        \Big]  \,. \label{eq:Omega_zwischen}
 \end{align}
Finally, we introduce the density of states of the spectrum of $H$ (i.e., the
Hamiltonian in the absence of diquark pairing) as
 \begin{equation} \label{eq:densityofstates}
        \rho(E) = \dfrac{1}{V} \sum\limits_\lambda \delta(E-E_\lambda)\,, 
 \end{equation}
in order to rewrite the sum over $\lambda$ in terms of an integral over the energy $E$. 
This yields
\begin{align}
        \Omega_\mathrm{kin} = 
         -\dfrac{1}{2} \sum\limits_{i=\pm}  \, \int\limits_{-\infty}^\infty dE \, \rho(E)
         \Big[ 
         &\epsilon_{i} + 2T \, \textrm{ln} \, \Big( 1 + \mathrm{e}^{-\epsilon_{i}/T}\Big) \notag \\
         + &\frac{\epsilon^{(0)} _{i}}{2} + T \, \textrm{ln} \, \Big( 1 + \mathrm{e}^{-\epsilon^{(0)} _{i}/T}\Big)
        \Big] \, ,
\end{align}
with
\begin{equation}
   \epsilon_\pm= \sqrt{(E \mp \mu)^2 + |\Delta|^2}
\qquad \text{and} \qquad
    \epsilon_\pm^{(0)}  = |E \mp \mu|  \,.     
\label{eq:epsilon_pm0}
\end{equation}
Since the original spectrum and, as a consequence, $\rho(E)$ are symmetric around zero, we can replace
$\frac{1}{2} \int_{-\infty}^\infty dE$ by $\int_{0}^\infty dE$.
Our final result for the grand potential is then given by
 \begin{align} 
        \Omega          
        = 
         -\sum\limits_{i=\pm}  \, \int\limits_0^\infty dE \, \rho(E)
         \Big[ 
         &\epsilon_i + 2T \, \textrm{ln} \, \Big( 1 + \mathrm{e}^{-\epsilon_i/T}\Big)      
          \notag \\
          + &\frac{\epsilon^{(0)} _{i}}{2} + T \, \textrm{ln} \, \Big( 1 + \mathrm{e}^{-\epsilon^{(0)} _{i}/T}\Big)                  
         \Big]
         \nonumber \\         
        + \dfrac{1}{4 G} \frac{1}{V} \int d^3 \mathbf{x} \big[ \sigma^2(\mathbf{x}) &+ \vec{\pi}^{\,2}(\mathbf{x}) \big] + \dfrac{|\Delta|^2}{4 G_\Delta}  \,.     \label{eq:Omega_final}     
 \end{align}
 \textcolor{black}{Note again that $\rho(E)$
 is the density of states for the case without diquark pairing, while the effects of the latter enter only
 through the energies $\epsilon_\pm$, see Eq.~(\ref{eq:epsilon_pm0}), and the last term in Eq.~(\ref{eq:Omega_final}).
 Hence, if $\rho(E)$ is known, the extension to include
 homogeneous 2SC condensates is straightforward.}

\subsection{Inhomogeneous chiral condensates}

We now evaluate the grand potential (\ref{eq:Omega_final}) for the CDW and RKC configurations. The respective densities of states have been found in Ref.\ \cite{Nickel:2009wj}. It is already known that, without diquark condensation, the RKC solution is preferred over the CDW one, and we do not expect that this will change when accounting for diquark condensation. Nevertheless, here we study both Ans\"atze, in order to confirm this expectation.

\subsubsection{Chiral density wave}

For the one-dimensional CDW the
Ansatz for the inhomogeneous chiral condensate
is given as
\begin{align}\label{eq:CDW}
    \sigma(z) + i \gamma_5 \vec{\pi}(z)\cdot\vec{\tau} &= - M \, \textrm{e}^{i \gamma_5 \tau_3 q z} \, ,
\end{align}
with an amplitude $M$ and the wave number 
$q$ of the CDW. Without loss of
generality, we have chosen the one-dimensional modulation to align with the $z$-axis. In the case $q=0$, $M$ is equal to the constituent quark mass.

The eigenvalues
$E_\lambda$ for the CDW without a color-superconducting condensate have been determined in Refs.\ \cite{Dautry:1979bk,Kutschera:1990xk}.
Inserting these into Eqs.\ (\ref{eq:E_gapped}),
(\ref{eq:E_ungapped}) yields the respective eigenvalues in the presence of a color-superconducting condensate, which are
identical to those found in Ref.\ \cite{Sadzikowski:2002iy}.

The density of states (\ref{eq:densityofstates})
for the CDW has been found in 
Ref.\ \cite{Nickel:2009wj},
\begin{widetext}
    \begin{align}\label{eq:CDWdos}
    \rho_{\textrm{CDW}}(E) = \dfrac{N_f E}{2 \pi^2} & \bigg[ \theta(E - q - M) \sqrt{(E-q)^2 - M^2} \notag \\ 
    &+ \theta(E - q + M) \, \theta(E + q - M) \sqrt{(E+q)^2 - M^2}  \notag \\
    &+ \theta(q - M - E) \left( \sqrt{(E+q)^2 - M^2} - \sqrt{(E-q)^2 - M^2} \right) \bigg] \, .
    \end{align}
\end{widetext}  
With Eq.\ (\ref{eq:CDW}), the grand potential (\ref{eq:Omega_final}) reads
	\begin{align}\label{eq:potCDW}
        \Omega_{\textrm{CDW}} &= -\sum_{i=\pm} \int_{0}^{\infty} dE \, \rho_{\textrm{CDW}}(E) \,  \Bigg[ \epsilon_{i} + \dfrac{\epsilon_{i}^{(0)}}{2}  \notag \\
        &+  2 T \, \textrm{ln} \left( 1 + \textrm{e}^{-\epsilon_{i}/T} \right) + T \, \textrm{ln} \left( 1 + \textrm{e}^{-\epsilon_{i}^{(0)}/T} \right)  \Bigg] \notag \\
        &+ \dfrac{M^2}{4 G} + \dfrac{|\Delta|^2}{4 G_{\Delta}} \, .
    \end{align}

\subsubsection{Real-kink crystal}

For the one-dimensional RKC, the pion
field is set to zero and the Ansatz for the
sigma field reads
\begin{equation}
    \sigma(z) \equiv \nu D \dfrac{\textrm{sn}(Dz|\nu)\, \textrm{cn}(Dz|\nu)}{ \textrm{dn}(Dz|\nu) } \,, 
\end{equation}
where \textrm{sn}, \textrm{cn}, \textrm{dn} are Jacobi elliptic functions. The parameter $\nu$ determines the shape of the condensate: For $\nu \rightarrow 1$ the Ansatz becomes 
$D  \tanh (D z)$, i.e., a kink-like soliton
of amplitude $D$ and width $1/D$. For $\nu \rightarrow 0$ it becomes a sine of infinitesimal amplitude. 

The density of states for the RKC has already been computed in Ref.\ \cite{Nickel:2009wj},
    \begin{widetext}
    \begin{align}
    \rho_{\textrm{RKC}}(E)  = \dfrac{N_f E D}{\pi^2} & \Bigg\lbrace \theta(\sqrt{\tilde{\nu}} D - E) \left[ \pmb{\textrm{E}}(\tilde{\theta}|\tilde{\nu}) + \left( \dfrac{\pmb{\textrm{E}}(\nu)}{\pmb{\textrm{K}}(\nu)} - 1\right) \pmb{\textrm{F}}(\tilde{\theta}|\tilde{\nu}) \right] \notag \\ 
    &+ \theta(E - \sqrt{\tilde{\nu}} D)\, \theta(D - E) \left[ \pmb{\textrm{E}}(\tilde{\nu}) + \left( \dfrac{\pmb{\textrm{E}}(\nu)}{\pmb{\textrm{K}}(\nu)} - 1\right) \pmb{\textrm{K}}(\tilde{\nu}) \right]  \notag \\
    &+ \theta(E - D)\left[ \pmb{\textrm{E}}(\theta|\tilde{\nu}) + \left( \dfrac{\pmb{\textrm{E}}(\nu)}{\pmb{\textrm{K}}(\nu)} - 1\right) \pmb{\textrm{F}}(\theta|\tilde{\nu}) + \dfrac{\sqrt{(E^2 - D^2)(E^2 - \tilde{\nu}D^2)}}{E D}\right]  \Bigg\rbrace \, \, ,
    \end{align}
\end{widetext}
where \pmb{K}$(\cdot)$ are the complete and \pmb{F}$(\cdot|\cdot)$ the incomplete elliptic integrals of the first kind,
respectively, 
while \pmb{E}($\cdot$) are the complete and \pmb{E}($\cdot | \cdot$) the incomplete elliptic integral of the second kind. We followed the notational convention of Ref.\ 
\cite{Nickel:2009wj},
where $\tilde{\nu} = 1 - \nu$, $\tilde{\theta} = \textrm{arcsin}[E/(\sqrt{\tilde{\nu}}D)]$, and  $\theta = \textrm{arcsin}(D/E)$.

The grand potential for the RKC Ansatz is found to be
	\begin{align}\label{eq:potSol}
        \Omega_{\textrm{RKC}} &= -\sum_{i=\pm} \int_{0}^{\infty} dE \, \rho_{\textrm{RKC}} (E) \,  \Bigg[ \epsilon_{i} + \dfrac{\epsilon_{i}^{(0)}}{2}  \notag \\
        &+  2 T \, \textrm{ln} \left( 1 + \textrm{e}^{-\epsilon_{i}/T} \right) + T \, \textrm{ln} \left( 1 + \textrm{e}^{-\epsilon_{i}^{(0)}/T} \right)  \Bigg]  \notag \\
        &+ \dfrac{M^2}{4 G} + \dfrac{|\Delta|^2}{4 G_{\Delta}}\, ,
    \end{align}
where we introduced the 
average \textcolor{black}{squared} amplitude of the RKC,     
\begin{equation}
        M^2 \equiv \frac{1}{L} \int_{0}^{L} dz\,  |\sigma(z)|^2 \, ,
    \end{equation}
with the period $L \equiv 4\, \textrm{\pmb{K}}(\nu)/D$ of the RKC.
For later purposes, we also define the effective wave number of the RKC as $q \equiv 2\pi/L$.

\subsubsection{Regularization and model parameters}
\label{sec:pauli}

Due to scattering of the quarks with the crystal the quasi-particle energies in the inhomogeneous phase cannot be labelled by a conserved three-momentum. Thus the commonly used 3-dimensional momentum cutoff regularization is not suitable here. 
In Refs.~\cite{Sadzikowski:2002iy,Sadzikowski:2006jq} this problem was circumvented by 
isolating a wave-number independent part of the 
grand potential and applying a momentum cutoff only
to this part. 
It turns out, however, that this regularization 
procedure leads to artifacts. 
For instance, even in the absence of diquark condensates,
the Lifshitz point does not coincide with the tricritical point, 
in contradiction to the general Ginzburg-Landau
result of Ref.~\cite{Nickel:2009ke}.
We therefore follow Ref.~\cite{Nowakowski:2016dwu} and apply a Pauli-Villars regularization scheme,
which acts on the energy spectrum and is commonly
used in the context of inhomogeneous phases.
In our case this
amounts to a replacement of
    \begin{align}
        \epsilon_{\pm} &= \sqrt{(E \pm \mu)^2 + |\Delta|^2} \notag \\
         &\rightarrow \sum_j c_j \sqrt{\left(\sqrt{E^2 + j \Lambda^2} \pm \mu\right)^2 + |\Delta|^2} \, 
    \end{align}
in the temperature-independent part of the thermodynamic potential.
    Following Ref.~\cite{Nickel:2009wj}, we take three regulators 
    with the coeffcients $c_0 = -c_3 = 1, c_1 = -c_2 = -3$,
    and fix the cutoff parameter $\Lambda$ together
    with the coupling constant $G$ by fitting \cite{Klevansky:1992qe} the
    constituent quark mass in vacuum to $M = 300$~MeV and the vacuum pion decay constant
    to its (approximate) value in the chiral limit, $f_\pi = 88$~MeV,
    which yields $\Lambda = 757.048$~MeV    
    and $G\Lambda^2 = 6.002$.
    
    In principle, the diquark coupling $G_\Delta$ can be obtained from a Fierz transformation of the interaction part of the Lagrangian (\ref{eq:Lagr_NJL}).
    In a color-current interaction model based on one-gluon exchange, it is found to be $G_\Delta = 3/4 \, G$ \cite{Buballa:2003qv}. However, in our setup this value leads to diquark gap parameters of about 200 MeV, which is significantly larger than the value expected from perturbative QCD \cite{Rischke:2003mt} 
    and other model calculations \cite{Alford:1997zt,Rapp:1997zu,Schwarz:1999dj}.
Therefore, we will treat $G_\Delta$ as a free parameter and present phase diagrams for four different diquark couplings: $G_\Delta = 0$, $0.3 \,G$, $G/2$, and $3G/4$.

\section{Results}
\label{sec:results}

In this section we present the phase diagrams for the CDW and RKC Ansatz, respectively, and elucidate the resulting phase structure by comparing $\Omega_{\textrm{CDW}}$ and $\Omega_{\textrm{RKC}}$ with the grand potential in the absence of chiral symmetry breaking or color superconductivity. 

\begin{widetext}

\begin{figure}[H]
\centering
\begin{subfigure}{.5\textwidth}
  \centering
  \includegraphics[width=1\linewidth]{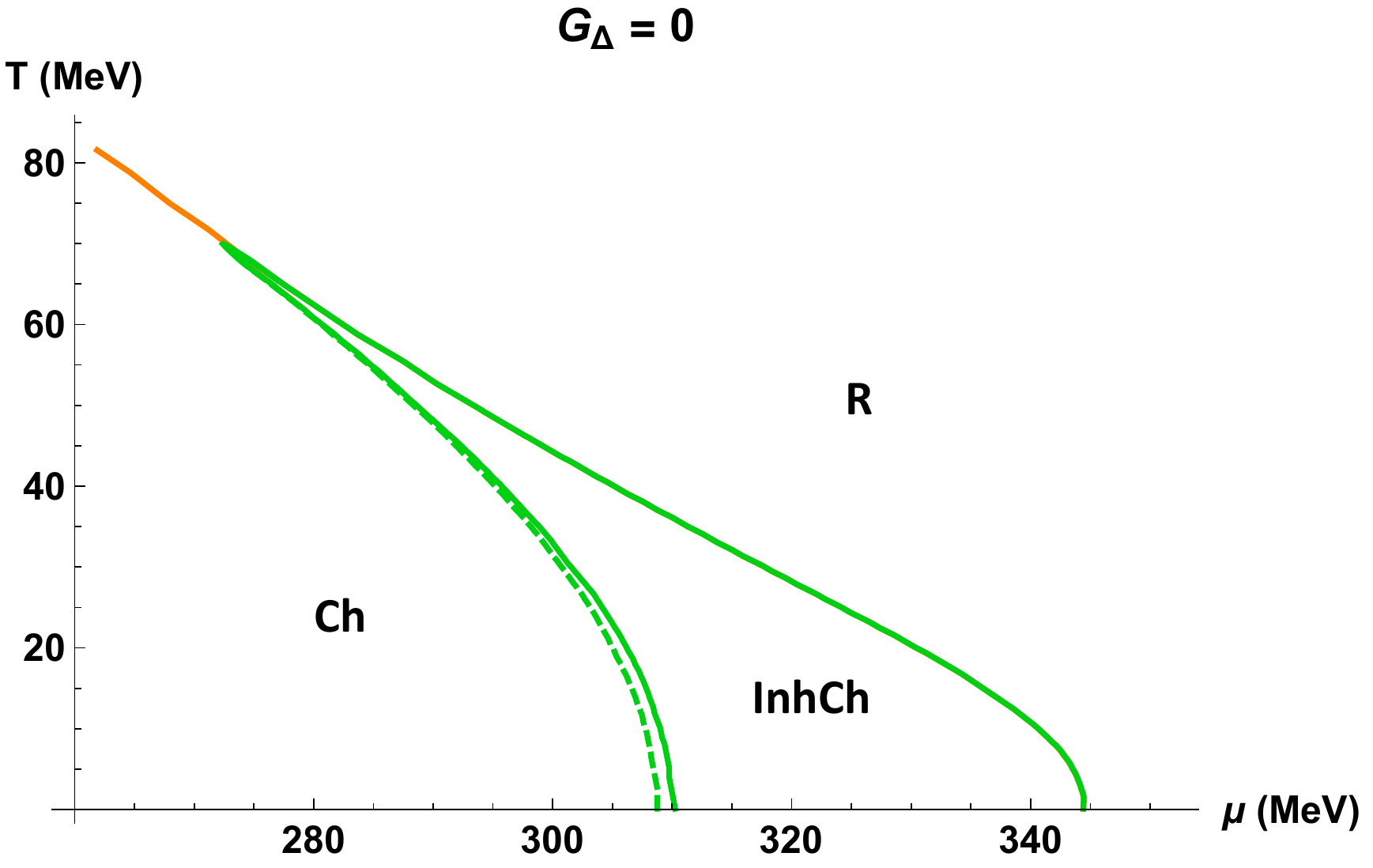}
  \label{fig:phaseBndsGd0}
\end{subfigure}%
\begin{subfigure}{.5\textwidth}
  \centering
  \includegraphics[width=1\linewidth]{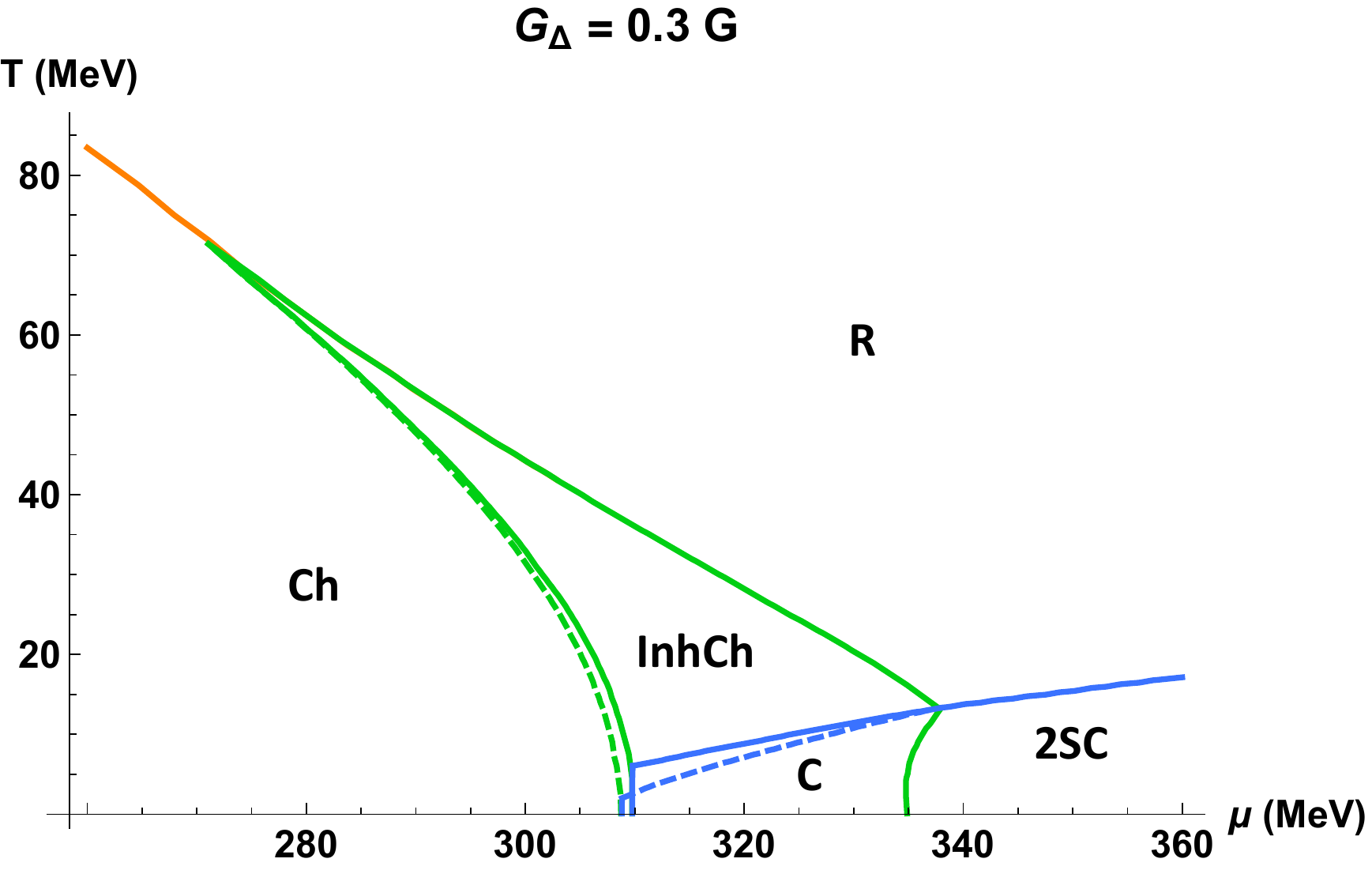}
  \label{fig:phaseBndsGd0.3}
\end{subfigure}

\centering
\begin{subfigure}{.5\textwidth}
  \centering
  \includegraphics[width=1\linewidth]{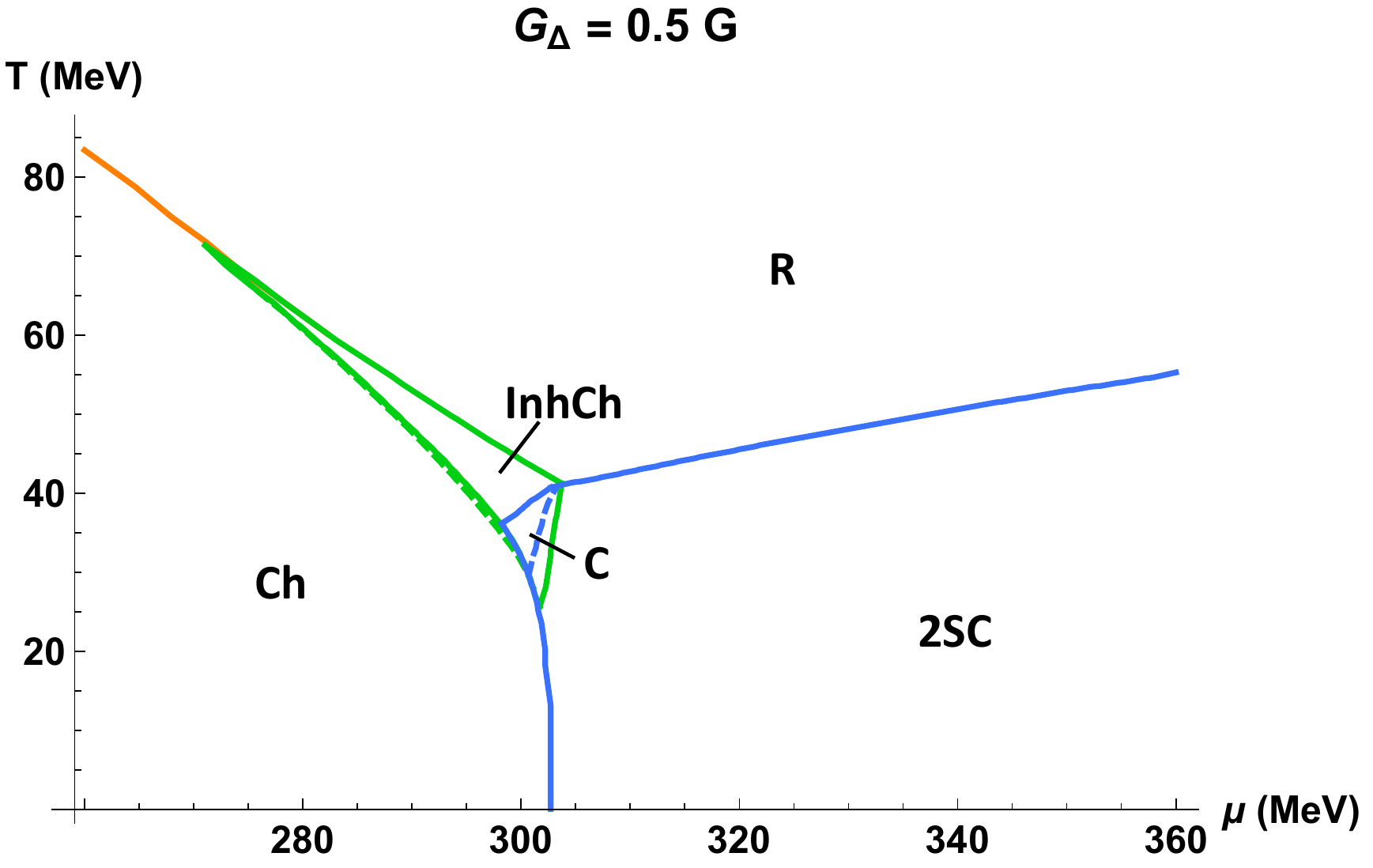}
  \label{fig:phaseBndsGd0.5}
\end{subfigure}%
\begin{subfigure}{.5\textwidth}
  \centering
  \includegraphics[width=1\linewidth]{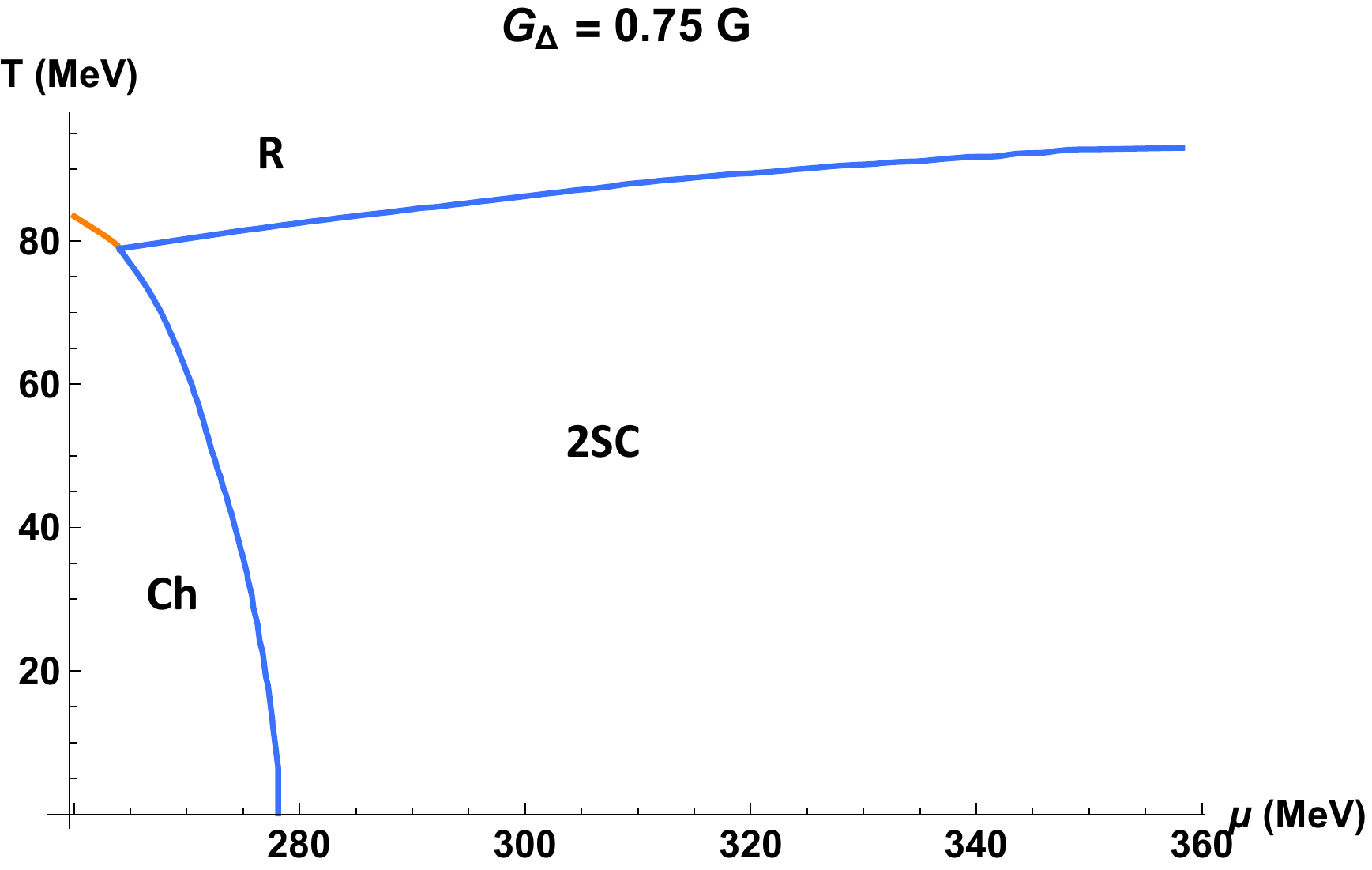}
  \label{fig:phaseBndsGd0.75}
\end{subfigure}
  \caption{The phase boundaries for different diquark couplings $G_\Delta = 0, 0.3 \, G,  G/2, 3G/4$. The solid lines correspond to the phase transition line for the CDW Ansatz, while the dashed lines are associated with the RKC Ansatz. 
  Where 
  only solid lines \textcolor{black}{are  visible,} the phase 
  boundaries for both
  Ans\"atze coincide.}\label{fig:phaseBoundaries}
\end{figure}
\end{widetext}

The phase diagrams are shown in Fig.\ \ref{fig:phaseBoundaries}.
We find five different phases:
\begin{itemize}
    \item The chiral symmetry-restored phase (R): $M = 0, \,q = 0, \,\Delta = 0$.
    \item The homogeneous chiral symmetry-broken phase (Ch): $M \ne 0, \, q = 0,\, \Delta = 0$.
     \item The inhomogeneous chiral phase (InhCh): $M \ne 0,\, q \ne 0,\, \Delta = 0$.
    \item The 2SC phase (2SC): $M = 0,\, q = 0, \,\Delta \ne 0$.
    \item The coexistence phase (C): $M \ne 0,\, q \ne 0, \,\Delta \ne 0$.
\end{itemize}

Regions with inhomogeneous chiral condensates are bounded by green lines and 
regions with
a nonvanishing 2SC condensate by blue lines. 
Solid lines represent the phase boundaries for the CDW Ansatz, while dashed lines correspond to the RKC Ansatz. The orange lines separate the homogeneous chiral symmetry-broken phase from the restored phase.
Note that in all cases where an inhomogeneous phase occurs, the tricritical point coincides with the Lifshitz point, in agreement with the results of Ref.\ \cite{Nickel:2009ke}.

In the upper left panel of Fig.\ \ref{fig:phaseBoundaries} we present the phase diagram in the case $G_\Delta = 0$.  For the RKC Ansatz, the inhomogeneous region extends to slightly smaller $\mu$
as compared to the CDW Ansatz. On the other hand, the boundary between the inhomogeneous and the chiral symmetry-restored phase
coincides for both Ans\"atze. 
At this boundary, the transition is of second order. These features are in agreement with previous studies, see, e.g., Ref.\ \cite{Carignano:2011gr}.

For $G_\Delta = 0.3 \, G$, cf.\ upper right panel of Fig.~\ref{fig:phaseBoundaries}, we find for both Ans\"atze a phase at low temperatures where an inhomogeneous chiral condensate and a diquark gap coexist (coexistence phase). Again, for the RKC this phase sets in for slightly smaller $\mu$ compared to the CDW. On the other hand, the
transition temperature between the coexistence and the
purely inhomogeneous phase is somewhat smaller for the RKC Ansatz compared to the CDW Ansatz. 

In this context, another interesting finding is that the coexistence phase sets in at smaller values of the chemical potential than a pure 2SC phase without inhomogeneous chiral condensate. This effect is rather small, and it is analogous to the
fact that the phase transition between the homogeneous chiral symmetry-broken phase to the inhomogeneous phase
occurs at slightly smaller $\mu$ than
the first-order phase transition between
the homogeneous chiral symmetry-broken 
and the restored phase \cite{Nakano:2004cd}.

This is best seen in Fig.\ \ref{fig:OmegaDiff}, where we compare the grand potentials for the different homogeneous and inhomogeneous phases. Each line represents a local extremum associated with the phase 
denoted in the legend of the figure. Intersections of two lines correspond to first-order phase transitions between the two respective phases, while converging lines characterize two converging extrema, corresponding to a second-order phase transition.

This figure yields several further insights. First, we can verify the result of Ref.\ \cite{Carignano:2011gr}. There the authors found that, while for the CDW Ansatz we have a first-order phase transition between the chiral symmetry-broken phase to the inhomogeneous chiral phase, this becomes a second-order transition when employing the RKC Ansatz. Secondly, this qualitative finding remains valid when including a 2SC phase, but here the second-order transition is to the coexistence phase instead of to the inhomogeneous chiral phase.

For further illustration we also provide three-dimensional plots in Fig.\ \ref{fig:3DPlot}, showing the values of the order parameters over the $T-\mu$ plane. While both plots for the CDW and the RKC Ansatz look very similar, one can see the different behavior for the wave number close to the Ch-C phase boundary line. 
For the CDW Ansatz, the wave number jumps from zero to a finite value, corresponding to the first-order transition, while for the RKC Ansatz, it increases \textcolor{black}{very steeply but} continuously at the second-order transition.
Upon closer inspection, one can also see a small discontinuous onset of the diquark gap for the CDW Ansatz, while the onset becomes continuous for the RKC case.

\textcolor{black}{
Increasing the coupling $G_\Delta$, the 2SC grand potential decreases while those of the phases 
without diquark pairing remain unchanged. 
As a consequence the inhomogeneous chiral phase becomes less favored compared to the 2SC phase, in particular at low temperatures where the gap parameter is largest.
This can be seen when we}
return to the discussion of Fig.\  \ref{fig:phaseBoundaries}. For $G_\Delta = G/2$, lower left panel, the maximum value of the diquark gap is $\sim 100$ MeV.
Here, the 2SC phase 
\textcolor{black}{is already dominant at low temperature,}
pushing the coexistence phase to higher temperatures. The latter is again smaller for the RKC Ansatz than for the CDW Ansatz. Also, the inhomogeneous chiral phase shrinks significantly.

Finally, in the lower right panel of Fig.\  \ref{fig:phaseBoundaries} we increased the diquark coupling to $G_\Delta = 3G/4$. For this parameter set, neither an inhomogeneous chiral phase nor a coexistence phase is found and therefore the phase boundaries for both the CDW and the RKC Ansatz exactly align. Note that this is in contrast to Ref.\ \cite{Sadzikowski:2006jq}, where the inhomogeneous chiral phase and the coexistence phase were also found in the case $G_\Delta = 3G/4$. We perceive this to be due to the different regularization scheme used in that work.

\begin{widetext}

\begin{figure}[H]
    \centering
    \includegraphics[width=0.99\textwidth]{
    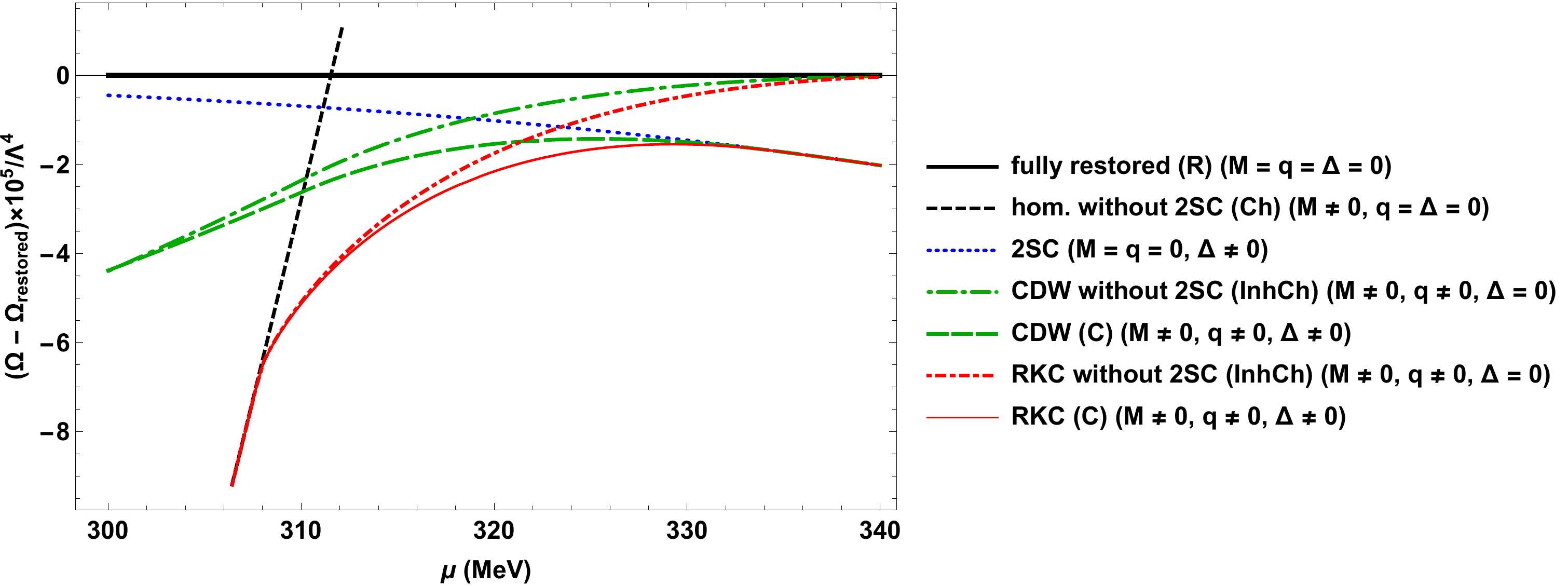}
    \caption{Comparison of the grand potentials for different scenarios at $T = 0$ MeV and for $G_\Delta = 0.3 \, G$. Two intersecting lines represent first-order transitions between the two respective phases, while converging lines characterize second-order phase transitions. For each case we subtract the potential of the fully restored ($M = q = \Delta = 0$) solution.}\label{fig:OmegaDiff}
\end{figure}

\begin{figure}[H]\label{fig:3DGd0.3}
\includegraphics[width=1\textwidth]{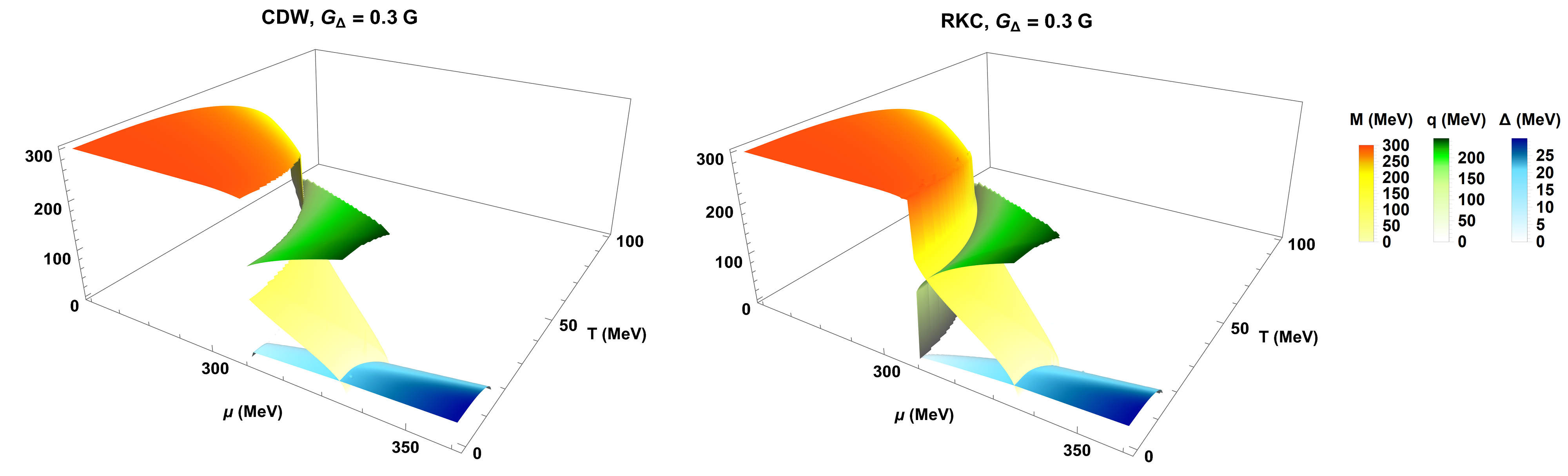}
  \caption{A detailed view of the order parameters for the CDW and the RKC Ansatz for $G_\Delta = 0.3 \, G$.}\label{fig:3DPlot}
\end{figure}

\end{widetext}

\section{Conclusions}
\label{sec:discussion}

In this paper we have investigated the phase diagram of the chirally symmetric two-flavor NJL model with two-flavor color superconductivity. The goal was to study the competition of inhomogeneous chiral phases and 2SC phases. 

To this end, we derived the grand potential for the chirally symmetric two-flavor NJL model with 2SC diquarks for a generic inhomogeneous condensate, formulated as energy integral over a generic density of states. In this formulation we were able to employ already known densities of states for nonuniform chiral condensates, namely the CDW and the RKC Ansatz. We investigated both Ans\"atze because in previous similar studies as in Refs.\ \cite{Sadzikowski:2002iy,Sadzikowski:2006jq}, the CDW has been used, but 
since then it has been found that the RKC is 
\textcolor{black}{in most cases}
energetically preferred over the CDW Ansatz 
\textcolor{black}{
\cite{Nickel:2009wj, Carignano:2010ac}, at least for
vanishing vector interactions \cite{Carignano:2018hvn}}.

We have found that for some diquark couplings the inhomogeneous chiral and the homogeneous 2SC phases are not necessarily excluding each other, but that these two phases may also
coexist. For the CDW Ansatz, such a coexistence
phase was already found in Refs.\ \cite{Sadzikowski:2002iy,Sadzikowski:2006jq}.
Here, we confirm a coexistence phase also for the RKC Ansatz. 

A further result of our study is that the specific shape of the inhomogeneous chiral condensate has an impact on the size of the inhomogeneous region as well as the size of the coexistence region. Since the RKC Ansatz is
energetically preferred over the CDW Ansatz, the inhomogeneous region is larger for the former. On the other hand, this also implies that the coexistence region becomes smaller for the RKC Ansatz.
Another interesting observation is that, in the case where we allow for a coexistence phase, the onset of this phase, and thus of diquark condensation, at $T = 0$ occurs at slightly smaller $\mu$ compared to a scenario where no coexistence phase is allowed.

There are several ways to continue research in this direction.
The NJL model has already been studied in many variations, e.g., by adding further degrees of freedom, such as a vector-channel interactions and Polyakov-loop dynamics \cite{Fukushima:2008wg, Bratovic:2012qs, Carignano:2010ac,Carignano:2018hvn}, or by including finite quark-mass effects \cite{Buballa:2018hux} and considering isospin-asymmetric matter \cite{Nowakowski:2015ksa, Nowakowski:2016dwu}. All these extensions could in principle also be studied together with color superconductivity.

Another direction when studying inhomogeneous phases is the inclusion of bosonic fluctuations, which, with the exception of Ref.\ \cite{Lenz:2020bxk}, has so far been largely ignored 
\textcolor{black}{[see, however, Refs.\
\cite{Lee:2015bva, Hidaka:2015xza, Yoshiike:2017kbx, Pisarski:2018bct, Pisarski:2020dnx}
for thorough discussions of fluctuation effects on inhomogeneous phases]}.
In a recent study \cite{Tripolt:2017zgc}, the authors performed a stability analysis in the QM model within the functional renormalization-group formalism \cite{Wetterich:1992yh}. In that study, the instability signals the formation of an inhomogeneous phase and is found in a region of the phase diagram where in mean-field studies inhomogeneous phases are typically found as well.
As a next step, we would like to extend this study by including two-flavor color superconductivity. It would be interesting to see if the results of the present work also hold when considering bosonic fluctuations.
Finally, we note that our results may have important implications for low-energy heavy-ion collisions, the structure of neutron stars, and the dynamics of binary neutron-star mergers.
\medskip 

\begin{acknowledgments}
P.L.\ thanks N.\ Cichutek, L.\ Kiefer, A.S.\ Salek, and D.\ Kraatz for useful discussions.
The authors acknowledge support by the Deutsche Forschungsgemeinschaft (DFG, German Research Foundation) through the CRC-TR 211 ``Strong-interaction matter under extreme conditions''– project number 315477589 – TRR 211.

\end{acknowledgments}

\bibliography{NJL2SC}{}

\begin{thebibliography}{43}%
\makeatletter
\providecommand \@ifxundefined [1]{%
 \@ifx{#1\undefined}
}%
\providecommand \@ifnum [1]{%
 \ifnum #1\expandafter \@firstoftwo
 \else \expandafter \@secondoftwo
 \fi
}%
\providecommand \@ifx [1]{%
 \ifx #1\expandafter \@firstoftwo
 \else \expandafter \@secondoftwo
 \fi
}%
\providecommand \natexlab [1]{#1}%
\providecommand \enquote  [1]{``#1''}%
\providecommand \bibnamefont  [1]{#1}%
\providecommand \bibfnamefont [1]{#1}%
\providecommand \citenamefont [1]{#1}%
\providecommand \href@noop [0]{\@secondoftwo}%
\providecommand \href [0]{\begingroup \@sanitize@url \@href}%
\providecommand \@href[1]{\@@startlink{#1}\@@href}%
\providecommand \@@href[1]{\endgroup#1\@@endlink}%
\providecommand \@sanitize@url [0]{\catcode `\\12\catcode `\$12\catcode
  `\&12\catcode `\#12\catcode `\^12\catcode `\_12\catcode `\%12\relax}%
\providecommand \@@startlink[1]{}%
\providecommand \@@endlink[0]{}%
\providecommand \url  [0]{\begingroup\@sanitize@url \@url }%
\providecommand \@url [1]{\endgroup\@href {#1}{\urlprefix }}%
\providecommand \urlprefix  [0]{URL }%
\providecommand \Eprint [0]{\href }%
\providecommand \doibase [0]{http://dx.doi.org/}%
\providecommand \selectlanguage [0]{\@gobble}%
\providecommand \bibinfo  [0]{\@secondoftwo}%
\providecommand \bibfield  [0]{\@secondoftwo}%
\providecommand \translation [1]{[#1]}%
\providecommand \BibitemOpen [0]{}%
\providecommand \bibitemStop [0]{}%
\providecommand \bibitemNoStop [0]{.\EOS\space}%
\providecommand \EOS [0]{\spacefactor3000\relax}%
\providecommand \BibitemShut  [1]{\csname bibitem#1\endcsname}%
\let\auto@bib@innerbib\@empty
\bibitem [{\citenamefont {Sadzikowski}(2003)}]{Sadzikowski:2002iy}%
  \BibitemOpen
  \bibfield  {author} {\bibinfo {author} {\bibfnamefont {M.}~\bibnamefont
  {Sadzikowski}},\ }\href {\doibase 10.1016/S0370-2693(02)03188-X} {\bibfield
  {journal} {\bibinfo  {journal} {Phys. Lett. B}\ }\textbf {\bibinfo {volume}
  {553}},\ \bibinfo {pages} {45} (\bibinfo {year} {2003})},\ \Eprint
  {http://arxiv.org/abs/hep-ph/0210065} {arXiv:hep-ph/0210065} \BibitemShut
  {NoStop}%
\bibitem [{\citenamefont {Sadzikowski}(2006)}]{Sadzikowski:2006jq}%
  \BibitemOpen
  \bibfield  {author} {\bibinfo {author} {\bibfnamefont {M.}~\bibnamefont
  {Sadzikowski}},\ }\href {\doibase 10.1016/j.physletb.2006.08.086} {\bibfield
  {journal} {\bibinfo  {journal} {Phys. Lett. B}\ }\textbf {\bibinfo {volume}
  {642}},\ \bibinfo {pages} {238} (\bibinfo {year} {2006})},\ \Eprint
  {http://arxiv.org/abs/hep-ph/0609186} {arXiv:hep-ph/0609186} \BibitemShut
  {NoStop}%
\bibitem [{\citenamefont {Bazavov}\ \emph {et~al.}(2019)\citenamefont {Bazavov}
  \emph {et~al.}}]{Bazavov:2018mes}%
  \BibitemOpen
  \bibfield  {author} {\bibinfo {author} {\bibfnamefont {A.}~\bibnamefont
  {Bazavov}} \emph {et~al.} (\bibinfo {collaboration} {HotQCD}),\ }\href
  {\doibase 10.1016/j.physletb.2019.05.013} {\bibfield  {journal} {\bibinfo
  {journal} {Phys. Lett. B}\ }\textbf {\bibinfo {volume} {795}},\ \bibinfo
  {pages} {15} (\bibinfo {year} {2019})},\ \Eprint
  {http://arxiv.org/abs/1812.08235} {arXiv:1812.08235 [hep-lat]} \BibitemShut
  {NoStop}%
\bibitem [{\citenamefont {Gao}\ and\ \citenamefont
  {Pawlowski}(2020)}]{Gao:2020fbl}%
  \BibitemOpen
  \bibfield  {author} {\bibinfo {author} {\bibfnamefont {F.}~\bibnamefont
  {Gao}}\ and\ \bibinfo {author} {\bibfnamefont {J.~M.}\ \bibnamefont
  {Pawlowski}},\ }\href@noop {} {\  (\bibinfo {year} {2020})},\ \Eprint
  {http://arxiv.org/abs/2010.13705} {arXiv:2010.13705 [hep-ph]} \BibitemShut
  {NoStop}%
\bibitem [{\citenamefont {Thies}(2006)}]{Thies:2006ti}%
  \BibitemOpen
  \bibfield  {author} {\bibinfo {author} {\bibfnamefont {M.}~\bibnamefont
  {Thies}},\ }\href {\doibase 10.1088/0305-4470/39/41/S04} {\bibfield
  {journal} {\bibinfo  {journal} {J. Phys. A}\ }\textbf {\bibinfo {volume}
  {39}},\ \bibinfo {pages} {12707} (\bibinfo {year} {2006})},\ \Eprint
  {http://arxiv.org/abs/hep-th/0601049} {arXiv:hep-th/0601049} \BibitemShut
  {NoStop}%
\bibitem [{\citenamefont {Sadzikowski}\ and\ \citenamefont
  {Broniowski}(2000)}]{Sadzikowski:2000ap}%
  \BibitemOpen
  \bibfield  {author} {\bibinfo {author} {\bibfnamefont {M.}~\bibnamefont
  {Sadzikowski}}\ and\ \bibinfo {author} {\bibfnamefont {W.}~\bibnamefont
  {Broniowski}},\ }\href {\doibase 10.1016/S0370-2693(00)00830-3} {\bibfield
  {journal} {\bibinfo  {journal} {Phys. Lett. B}\ }\textbf {\bibinfo {volume}
  {488}},\ \bibinfo {pages} {63} (\bibinfo {year} {2000})},\ \Eprint
  {http://arxiv.org/abs/hep-ph/0003282} {arXiv:hep-ph/0003282} \BibitemShut
  {NoStop}%
\bibitem [{\citenamefont {Nakano}\ and\ \citenamefont
  {Tatsumi}(2005)}]{Nakano:2004cd}%
  \BibitemOpen
  \bibfield  {author} {\bibinfo {author} {\bibfnamefont {E.}~\bibnamefont
  {Nakano}}\ and\ \bibinfo {author} {\bibfnamefont {T.}~\bibnamefont
  {Tatsumi}},\ }\href {\doibase 10.1103/PhysRevD.71.114006} {\bibfield
  {journal} {\bibinfo  {journal} {Phys. Rev. D}\ }\textbf {\bibinfo {volume}
  {71}},\ \bibinfo {pages} {114006} (\bibinfo {year} {2005})},\ \Eprint
  {http://arxiv.org/abs/hep-ph/0411350} {arXiv:hep-ph/0411350} \BibitemShut
  {NoStop}%
\bibitem [{\citenamefont {Nickel}(2009{\natexlab{a}})}]{Nickel:2009wj}%
  \BibitemOpen
  \bibfield  {author} {\bibinfo {author} {\bibfnamefont {D.}~\bibnamefont
  {Nickel}},\ }\href {\doibase 10.1103/PhysRevD.80.074025} {\bibfield
  {journal} {\bibinfo  {journal} {Phys. Rev. D}\ }\textbf {\bibinfo {volume}
  {80}},\ \bibinfo {pages} {074025} (\bibinfo {year} {2009}{\natexlab{a}})},\
  \Eprint {http://arxiv.org/abs/0906.5295} {arXiv:0906.5295 [hep-ph]}
  \BibitemShut {NoStop}%
\bibitem [{\citenamefont {Carignano}\ \emph {et~al.}(2014)\citenamefont
  {Carignano}, \citenamefont {Buballa},\ and\ \citenamefont
  {Schaefer}}]{Carignano:2014jla}%
  \BibitemOpen
  \bibfield  {author} {\bibinfo {author} {\bibfnamefont {S.}~\bibnamefont
  {Carignano}}, \bibinfo {author} {\bibfnamefont {M.}~\bibnamefont {Buballa}},
  \ and\ \bibinfo {author} {\bibfnamefont {B.-J.}\ \bibnamefont {Schaefer}},\
  }\href {\doibase 10.1103/PhysRevD.90.014033} {\bibfield  {journal} {\bibinfo
  {journal} {Phys. Rev. D}\ }\textbf {\bibinfo {volume} {90}},\ \bibinfo
  {pages} {014033} (\bibinfo {year} {2014})},\ \Eprint
  {http://arxiv.org/abs/1404.0057} {arXiv:1404.0057 [hep-ph]} \BibitemShut
  {NoStop}%
\bibitem [{\citenamefont {Adhikari}\ \emph {et~al.}(2017)\citenamefont
  {Adhikari}, \citenamefont {Andersen},\ and\ \citenamefont
  {Kneschke}}]{Adhikari:2017ydi}%
  \BibitemOpen
  \bibfield  {author} {\bibinfo {author} {\bibfnamefont {P.}~\bibnamefont
  {Adhikari}}, \bibinfo {author} {\bibfnamefont {J.~O.}\ \bibnamefont
  {Andersen}}, \ and\ \bibinfo {author} {\bibfnamefont {P.}~\bibnamefont
  {Kneschke}},\ }\href {\doibase 10.1103/PhysRevD.96.016013} {\bibfield
  {journal} {\bibinfo  {journal} {Phys. Rev. D}\ }\textbf {\bibinfo {volume}
  {96}},\ \bibinfo {pages} {016013} (\bibinfo {year} {2017})},\ \bibinfo {note}
  {[Erratum: Phys.Rev.D 98, 099902 (2018)]},\ \Eprint
  {http://arxiv.org/abs/1702.01324} {arXiv:1702.01324 [hep-ph]} \BibitemShut
  {NoStop}%
\bibitem [{\citenamefont {Heinz}\ \emph {et~al.}(2015)\citenamefont {Heinz},
  \citenamefont {Giacosa},\ and\ \citenamefont {Rischke}}]{Heinz:2013hza}%
  \BibitemOpen
  \bibfield  {author} {\bibinfo {author} {\bibfnamefont {A.}~\bibnamefont
  {Heinz}}, \bibinfo {author} {\bibfnamefont {F.}~\bibnamefont {Giacosa}}, \
  and\ \bibinfo {author} {\bibfnamefont {D.~H.}\ \bibnamefont {Rischke}},\
  }\href {\doibase 10.1016/j.nuclphysa.2014.09.027} {\bibfield  {journal}
  {\bibinfo  {journal} {Nucl. Phys. A}\ }\textbf {\bibinfo {volume} {933}},\
  \bibinfo {pages} {34} (\bibinfo {year} {2015})},\ \Eprint
  {http://arxiv.org/abs/1312.3244} {arXiv:1312.3244 [nucl-th]} \BibitemShut
  {NoStop}%
\bibitem [{\citenamefont {Buballa}\ and\ \citenamefont
  {Carignano}(2015)}]{Buballa:2014tba}%
  \BibitemOpen
  \bibfield  {author} {\bibinfo {author} {\bibfnamefont {M.}~\bibnamefont
  {Buballa}}\ and\ \bibinfo {author} {\bibfnamefont {S.}~\bibnamefont
  {Carignano}},\ }\href {\doibase 10.1016/j.ppnp.2014.11.001} {\bibfield
  {journal} {\bibinfo  {journal} {Prog. Part. Nucl. Phys.}\ }\textbf {\bibinfo
  {volume} {81}},\ \bibinfo {pages} {39} (\bibinfo {year} {2015})},\ \Eprint
  {http://arxiv.org/abs/1406.1367} {arXiv:1406.1367 [hep-ph]} \BibitemShut
  {NoStop}%
\bibitem [{\citenamefont {Abuki}\ \emph {et~al.}(2012)\citenamefont {Abuki},
  \citenamefont {Ishibashi},\ and\ \citenamefont {Suzuki}}]{Abuki:2011pf}%
  \BibitemOpen
  \bibfield  {author} {\bibinfo {author} {\bibfnamefont {H.}~\bibnamefont
  {Abuki}}, \bibinfo {author} {\bibfnamefont {D.}~\bibnamefont {Ishibashi}}, \
  and\ \bibinfo {author} {\bibfnamefont {K.}~\bibnamefont {Suzuki}},\ }\href
  {\doibase 10.1103/PhysRevD.85.074002} {\bibfield  {journal} {\bibinfo
  {journal} {Phys. Rev. D}\ }\textbf {\bibinfo {volume} {85}},\ \bibinfo
  {pages} {074002} (\bibinfo {year} {2012})},\ \Eprint
  {http://arxiv.org/abs/1109.1615} {arXiv:1109.1615 [hep-ph]} \BibitemShut
  {NoStop}%
\bibitem [{\citenamefont {Nickel}(2009{\natexlab{b}})}]{Nickel:2009ke}%
  \BibitemOpen
  \bibfield  {author} {\bibinfo {author} {\bibfnamefont {D.}~\bibnamefont
  {Nickel}},\ }\href {\doibase 10.1103/PhysRevLett.103.072301} {\bibfield
  {journal} {\bibinfo  {journal} {Phys. Rev. Lett.}\ }\textbf {\bibinfo
  {volume} {103}},\ \bibinfo {pages} {072301} (\bibinfo {year}
  {2009}{\natexlab{b}})},\ \Eprint {http://arxiv.org/abs/0902.1778}
  {arXiv:0902.1778 [hep-ph]} \BibitemShut {NoStop}%
\bibitem [{\citenamefont {Heinz}\ \emph {et~al.}(2016)\citenamefont {Heinz},
  \citenamefont {Giacosa}, \citenamefont {Wagner},\ and\ \citenamefont
  {Rischke}}]{Heinz:2015lua}%
  \BibitemOpen
  \bibfield  {author} {\bibinfo {author} {\bibfnamefont {A.}~\bibnamefont
  {Heinz}}, \bibinfo {author} {\bibfnamefont {F.}~\bibnamefont {Giacosa}},
  \bibinfo {author} {\bibfnamefont {M.}~\bibnamefont {Wagner}}, \ and\ \bibinfo
  {author} {\bibfnamefont {D.~H.}\ \bibnamefont {Rischke}},\ }\href {\doibase
  10.1103/PhysRevD.93.014007} {\bibfield  {journal} {\bibinfo  {journal} {Phys.
  Rev. D}\ }\textbf {\bibinfo {volume} {93}},\ \bibinfo {pages} {014007}
  (\bibinfo {year} {2016})},\ \Eprint {http://arxiv.org/abs/1508.06057}
  {arXiv:1508.06057 [hep-ph]} \BibitemShut {NoStop}%
\bibitem [{\citenamefont {Lenz}\ \emph {et~al.}(2020)\citenamefont {Lenz},
  \citenamefont {Pannullo}, \citenamefont {Wagner}, \citenamefont
  {Wellegehausen},\ and\ \citenamefont {Wipf}}]{Lenz:2020bxk}%
  \BibitemOpen
  \bibfield  {author} {\bibinfo {author} {\bibfnamefont {J.}~\bibnamefont
  {Lenz}}, \bibinfo {author} {\bibfnamefont {L.}~\bibnamefont {Pannullo}},
  \bibinfo {author} {\bibfnamefont {M.}~\bibnamefont {Wagner}}, \bibinfo
  {author} {\bibfnamefont {B.}~\bibnamefont {Wellegehausen}}, \ and\ \bibinfo
  {author} {\bibfnamefont {A.}~\bibnamefont {Wipf}},\ }\href {\doibase
  10.1103/PhysRevD.101.094512} {\bibfield  {journal} {\bibinfo  {journal}
  {Phys. Rev. D}\ }\textbf {\bibinfo {volume} {101}},\ \bibinfo {pages}
  {094512} (\bibinfo {year} {2020})},\ \Eprint
  {http://arxiv.org/abs/2004.00295} {arXiv:2004.00295 [hep-lat]} \BibitemShut
  {NoStop}%
\bibitem [{\citenamefont {Alford}\ \emph {et~al.}(1998)\citenamefont {Alford},
  \citenamefont {Rajagopal},\ and\ \citenamefont {Wilczek}}]{Alford:1997zt}%
  \BibitemOpen
  \bibfield  {author} {\bibinfo {author} {\bibfnamefont {M.~G.}\ \bibnamefont
  {Alford}}, \bibinfo {author} {\bibfnamefont {K.}~\bibnamefont {Rajagopal}}, \
  and\ \bibinfo {author} {\bibfnamefont {F.}~\bibnamefont {Wilczek}},\ }\href
  {\doibase 10.1016/S0370-2693(98)00051-3} {\bibfield  {journal} {\bibinfo
  {journal} {Phys. Lett. B}\ }\textbf {\bibinfo {volume} {422}},\ \bibinfo
  {pages} {247} (\bibinfo {year} {1998})},\ \Eprint
  {http://arxiv.org/abs/hep-ph/9711395} {arXiv:hep-ph/9711395} \BibitemShut
  {NoStop}%
\bibitem [{\citenamefont {Rischke}(2004)}]{Rischke:2003mt}%
  \BibitemOpen
  \bibfield  {author} {\bibinfo {author} {\bibfnamefont {D.}~\bibnamefont
  {Rischke}},\ }\href {\doibase 10.1016/j.ppnp.2003.09.002} {\bibfield
  {journal} {\bibinfo  {journal} {Progress in Particle and Nuclear Physics}\
  }\textbf {\bibinfo {volume} {52}},\ \bibinfo {pages} {197–296} (\bibinfo
  {year} {2004})}\BibitemShut {NoStop}%
\bibitem [{\citenamefont {Alford}\ \emph {et~al.}(2008)\citenamefont {Alford},
  \citenamefont {Schmitt}, \citenamefont {Rajagopal},\ and\ \citenamefont
  {Schäfer}}]{Alford:2007xm}%
  \BibitemOpen
  \bibfield  {author} {\bibinfo {author} {\bibfnamefont {M.~G.}\ \bibnamefont
  {Alford}}, \bibinfo {author} {\bibfnamefont {A.}~\bibnamefont {Schmitt}},
  \bibinfo {author} {\bibfnamefont {K.}~\bibnamefont {Rajagopal}}, \ and\
  \bibinfo {author} {\bibfnamefont {T.}~\bibnamefont {Schäfer}},\ }\href
  {\doibase 10.1103/RevModPhys.80.1455} {\bibfield  {journal} {\bibinfo
  {journal} {Rev. Mod. Phys.}\ }\textbf {\bibinfo {volume} {80}},\ \bibinfo
  {pages} {1455} (\bibinfo {year} {2008})},\ \Eprint
  {http://arxiv.org/abs/0709.4635} {arXiv:0709.4635 [hep-ph]} \BibitemShut
  {NoStop}%
\bibitem [{\citenamefont {Nowakowski}\ and\ \citenamefont
  {Carignano}(2016)}]{Nowakowski:2016dwu}%
  \BibitemOpen
  \bibfield  {author} {\bibinfo {author} {\bibfnamefont {D.}~\bibnamefont
  {Nowakowski}}\ and\ \bibinfo {author} {\bibfnamefont {S.}~\bibnamefont
  {Carignano}},\ }\href {\doibase 10.22323/1.262.0010} {\bibfield  {journal}
  {\bibinfo  {journal} {PoS}\ }\textbf {\bibinfo {volume} {MPCS2015}},\
  \bibinfo {pages} {010} (\bibinfo {year} {2016})},\ \Eprint
  {http://arxiv.org/abs/1602.04798} {arXiv:1602.04798 [hep-ph]} \BibitemShut
  {NoStop}%
\bibitem [{\citenamefont {Blaschke}\ \emph {et~al.}(2003)\citenamefont
  {Blaschke}, \citenamefont {Volkov},\ and\ \citenamefont
  {Yudichev}}]{Blaschke:2003cv}%
  \BibitemOpen
  \bibfield  {author} {\bibinfo {author} {\bibfnamefont {D.}~\bibnamefont
  {Blaschke}}, \bibinfo {author} {\bibfnamefont {M.}~\bibnamefont {Volkov}}, \
  and\ \bibinfo {author} {\bibfnamefont {V.}~\bibnamefont {Yudichev}},\ }\href
  {\doibase 10.1140/epja/i2003-10003-9} {\bibfield  {journal} {\bibinfo
  {journal} {Eur. Phys. J. A}\ }\textbf {\bibinfo {volume} {17}},\ \bibinfo
  {pages} {103} (\bibinfo {year} {2003})},\ \Eprint
  {http://arxiv.org/abs/hep-ph/0301065} {arXiv:hep-ph/0301065} \BibitemShut
  {NoStop}%
\bibitem [{\citenamefont {Schmitt}(2015)}]{Schmitt:2014eka}%
  \BibitemOpen
  \bibfield  {author} {\bibinfo {author} {\bibfnamefont {A.}~\bibnamefont
  {Schmitt}},\ }\href {\doibase 10.1007/978-3-319-07947-9} {\emph {\bibinfo
  {title} {{Introduction to Superfluidity}: {Field-theoretical approach and
  applications}}}},\ Vol.\ \bibinfo {volume} {888}\ (\bibinfo {year} {2015})\
  \Eprint {http://arxiv.org/abs/1404.1284} {arXiv:1404.1284 [hep-ph]}
  \BibitemShut {NoStop}%
\bibitem [{\citenamefont {Carignano}\ \emph {et~al.}(2010)\citenamefont
  {Carignano}, \citenamefont {Nickel},\ and\ \citenamefont
  {Buballa}}]{Carignano:2010ac}%
  \BibitemOpen
  \bibfield  {author} {\bibinfo {author} {\bibfnamefont {S.}~\bibnamefont
  {Carignano}}, \bibinfo {author} {\bibfnamefont {D.}~\bibnamefont {Nickel}}, \
  and\ \bibinfo {author} {\bibfnamefont {M.}~\bibnamefont {Buballa}},\ }\href
  {\doibase 10.1103/PhysRevD.82.054009} {\bibfield  {journal} {\bibinfo
  {journal} {Phys. Rev. D}\ }\textbf {\bibinfo {volume} {82}},\ \bibinfo
  {pages} {054009} (\bibinfo {year} {2010})},\ \Eprint
  {http://arxiv.org/abs/1007.1397} {arXiv:1007.1397 [hep-ph]} \BibitemShut
  {NoStop}%
\bibitem [{\citenamefont {Kapusta}\ and\ \citenamefont {Gale}(2011)}]{kapusta}%
  \BibitemOpen
  \bibfield  {author} {\bibinfo {author} {\bibfnamefont {J.}~\bibnamefont
  {Kapusta}}\ and\ \bibinfo {author} {\bibfnamefont {C.}~\bibnamefont {Gale}},\
  }\href {\doibase 10.1017/CBO9780511535130} {\emph {\bibinfo {title}
  {{Finite-temperature field theory: Principles and applications}}}},\
  Cambridge Monographs on Mathematical Physics\ (\bibinfo  {publisher}
  {Cambridge University Press},\ \bibinfo {year} {2011})\BibitemShut {NoStop}%
\bibitem [{\citenamefont {Dautry}\ and\ \citenamefont
  {Nyman}(1979)}]{Dautry:1979bk}%
  \BibitemOpen
  \bibfield  {author} {\bibinfo {author} {\bibfnamefont {F.}~\bibnamefont
  {Dautry}}\ and\ \bibinfo {author} {\bibfnamefont {E.}~\bibnamefont {Nyman}},\
  }\href {\doibase 10.1016/0375-9474(79)90518-9} {\bibfield  {journal}
  {\bibinfo  {journal} {Nucl. Phys. A}\ }\textbf {\bibinfo {volume} {319}},\
  \bibinfo {pages} {323} (\bibinfo {year} {1979})}\BibitemShut {NoStop}%
\bibitem [{\citenamefont {Kutschera}\ \emph {et~al.}(1990)\citenamefont
  {Kutschera}, \citenamefont {Broniowski},\ and\ \citenamefont
  {Kotlorz}}]{Kutschera:1990xk}%
  \BibitemOpen
  \bibfield  {author} {\bibinfo {author} {\bibfnamefont {M.}~\bibnamefont
  {Kutschera}}, \bibinfo {author} {\bibfnamefont {W.}~\bibnamefont
  {Broniowski}}, \ and\ \bibinfo {author} {\bibfnamefont {A.}~\bibnamefont
  {Kotlorz}},\ }\href {\doibase 10.1016/0375-9474(90)90128-9} {\bibfield
  {journal} {\bibinfo  {journal} {Nucl. Phys. A}\ }\textbf {\bibinfo {volume}
  {516}},\ \bibinfo {pages} {566} (\bibinfo {year} {1990})}\BibitemShut
  {NoStop}%
\bibitem [{\citenamefont {Klevansky}(1992)}]{Klevansky:1992qe}%
  \BibitemOpen
  \bibfield  {author} {\bibinfo {author} {\bibfnamefont {S.~P.}\ \bibnamefont
  {Klevansky}},\ }\href {\doibase 10.1103/RevModPhys.64.649} {\bibfield
  {journal} {\bibinfo  {journal} {Rev. Mod. Phys.}\ }\textbf {\bibinfo {volume}
  {64}},\ \bibinfo {pages} {649} (\bibinfo {year} {1992})}\BibitemShut
  {NoStop}%
\bibitem [{\citenamefont {Buballa}(2005)}]{Buballa:2003qv}%
  \BibitemOpen
  \bibfield  {author} {\bibinfo {author} {\bibfnamefont {M.}~\bibnamefont
  {Buballa}},\ }\href {\doibase 10.1016/j.physrep.2004.11.004} {\bibfield
  {journal} {\bibinfo  {journal} {Phys. Rept.}\ }\textbf {\bibinfo {volume}
  {407}},\ \bibinfo {pages} {205} (\bibinfo {year} {2005})},\ \Eprint
  {http://arxiv.org/abs/hep-ph/0402234} {arXiv:hep-ph/0402234} \BibitemShut
  {NoStop}%
\bibitem [{\citenamefont {Rapp}\ \emph {et~al.}(1998)\citenamefont {Rapp},
  \citenamefont {Sch\"afer}, \citenamefont {Shuryak},\ and\ \citenamefont
  {Velkovsky}}]{Rapp:1997zu}%
  \BibitemOpen
  \bibfield  {author} {\bibinfo {author} {\bibfnamefont {R.}~\bibnamefont
  {Rapp}}, \bibinfo {author} {\bibfnamefont {T.}~\bibnamefont {Sch\"afer}},
  \bibinfo {author} {\bibfnamefont {E.~V.}\ \bibnamefont {Shuryak}}, \ and\
  \bibinfo {author} {\bibfnamefont {M.}~\bibnamefont {Velkovsky}},\ }\href
  {\doibase 10.1103/PhysRevLett.81.53} {\bibfield  {journal} {\bibinfo
  {journal} {Phys. Rev. Lett.}\ }\textbf {\bibinfo {volume} {81}},\ \bibinfo
  {pages} {53} (\bibinfo {year} {1998})},\ \Eprint
  {http://arxiv.org/abs/hep-ph/9711396} {arXiv:hep-ph/9711396} \BibitemShut
  {NoStop}%
\bibitem [{\citenamefont {Schwarz}\ \emph {et~al.}(1999)\citenamefont
  {Schwarz}, \citenamefont {Klevansky},\ and\ \citenamefont
  {Papp}}]{Schwarz:1999dj}%
  \BibitemOpen
  \bibfield  {author} {\bibinfo {author} {\bibfnamefont {T.}~\bibnamefont
  {Schwarz}}, \bibinfo {author} {\bibfnamefont {S.}~\bibnamefont {Klevansky}},
  \ and\ \bibinfo {author} {\bibfnamefont {G.}~\bibnamefont {Papp}},\ }\href
  {\doibase 10.1103/PhysRevC.60.055205} {\bibfield  {journal} {\bibinfo
  {journal} {Phys. Rev. C}\ }\textbf {\bibinfo {volume} {60}},\ \bibinfo
  {pages} {055205} (\bibinfo {year} {1999})},\ \Eprint
  {http://arxiv.org/abs/nucl-th/9903048} {arXiv:nucl-th/9903048} \BibitemShut
  {NoStop}%
\bibitem [{\citenamefont {Carignano}\ and\ \citenamefont
  {Buballa}(2012)}]{Carignano:2011gr}%
  \BibitemOpen
  \bibfield  {author} {\bibinfo {author} {\bibfnamefont {S.}~\bibnamefont
  {Carignano}}\ and\ \bibinfo {author} {\bibfnamefont {M.}~\bibnamefont
  {Buballa}},\ }\href {\doibase 10.5506/APhysPolBSupp.5.641} {\bibfield
  {journal} {\bibinfo  {journal} {Acta Phys. Polon. Supp.}\ }\textbf {\bibinfo
  {volume} {5}},\ \bibinfo {pages} {641} (\bibinfo {year} {2012})},\ \Eprint
  {http://arxiv.org/abs/1111.4400} {arXiv:1111.4400 [hep-ph]} \BibitemShut
  {NoStop}%
\bibitem [{\citenamefont {Carignano}\ \emph {et~al.}(2018)\citenamefont
  {Carignano}, \citenamefont {Schramm},\ and\ \citenamefont
  {Buballa}}]{Carignano:2018hvn}%
  \BibitemOpen
  \bibfield  {author} {\bibinfo {author} {\bibfnamefont {S.}~\bibnamefont
  {Carignano}}, \bibinfo {author} {\bibfnamefont {M.}~\bibnamefont {Schramm}},
  \ and\ \bibinfo {author} {\bibfnamefont {M.}~\bibnamefont {Buballa}},\ }\href
  {\doibase 10.1103/PhysRevD.98.014033} {\bibfield  {journal} {\bibinfo
  {journal} {Phys. Rev. D}\ }\textbf {\bibinfo {volume} {98}},\ \bibinfo
  {pages} {014033} (\bibinfo {year} {2018})},\ \Eprint
  {http://arxiv.org/abs/1805.06203} {arXiv:1805.06203 [hep-ph]} \BibitemShut
  {NoStop}%
\bibitem [{\citenamefont {Fukushima}(2008)}]{Fukushima:2008wg}%
  \BibitemOpen
  \bibfield  {author} {\bibinfo {author} {\bibfnamefont {K.}~\bibnamefont
  {Fukushima}},\ }\href {\doibase 10.1103/PhysRevD.77.114028} {\bibfield
  {journal} {\bibinfo  {journal} {Phys. Rev. D}\ }\textbf {\bibinfo {volume}
  {77}},\ \bibinfo {pages} {114028} (\bibinfo {year} {2008})},\ \bibinfo {note}
  {[Erratum: Phys.Rev.D 78, 039902 (2008)]},\ \Eprint
  {http://arxiv.org/abs/0803.3318} {arXiv:0803.3318 [hep-ph]} \BibitemShut
  {NoStop}%
\bibitem [{\citenamefont {Bratovic}\ \emph {et~al.}(2013)\citenamefont
  {Bratovic}, \citenamefont {Hatsuda},\ and\ \citenamefont
  {Weise}}]{Bratovic:2012qs}%
  \BibitemOpen
  \bibfield  {author} {\bibinfo {author} {\bibfnamefont {N.~M.}\ \bibnamefont
  {Bratovic}}, \bibinfo {author} {\bibfnamefont {T.}~\bibnamefont {Hatsuda}}, \
  and\ \bibinfo {author} {\bibfnamefont {W.}~\bibnamefont {Weise}},\ }\href
  {\doibase 10.1016/j.physletb.2013.01.003} {\bibfield  {journal} {\bibinfo
  {journal} {Phys. Lett. B}\ }\textbf {\bibinfo {volume} {719}},\ \bibinfo
  {pages} {131} (\bibinfo {year} {2013})},\ \Eprint
  {http://arxiv.org/abs/1204.3788} {arXiv:1204.3788 [hep-ph]} \BibitemShut
  {NoStop}%
\bibitem [{\citenamefont {Buballa}\ and\ \citenamefont
  {Carignano}(2019)}]{Buballa:2018hux}%
  \BibitemOpen
  \bibfield  {author} {\bibinfo {author} {\bibfnamefont {M.}~\bibnamefont
  {Buballa}}\ and\ \bibinfo {author} {\bibfnamefont {S.}~\bibnamefont
  {Carignano}},\ }\href {\doibase 10.1016/j.physletb.2019.02.045} {\bibfield
  {journal} {\bibinfo  {journal} {Phys. Lett. B}\ }\textbf {\bibinfo {volume}
  {791}},\ \bibinfo {pages} {361} (\bibinfo {year} {2019})},\ \Eprint
  {http://arxiv.org/abs/1809.10066} {arXiv:1809.10066 [hep-ph]} \BibitemShut
  {NoStop}%
\bibitem [{\citenamefont {Nowakowski}\ \emph {et~al.}(2015)\citenamefont
  {Nowakowski}, \citenamefont {Buballa}, \citenamefont {Carignano},\ and\
  \citenamefont {Wambach}}]{Nowakowski:2015ksa}%
  \BibitemOpen
  \bibfield  {author} {\bibinfo {author} {\bibfnamefont {D.}~\bibnamefont
  {Nowakowski}}, \bibinfo {author} {\bibfnamefont {M.}~\bibnamefont {Buballa}},
  \bibinfo {author} {\bibfnamefont {S.}~\bibnamefont {Carignano}}, \ and\
  \bibinfo {author} {\bibfnamefont {J.}~\bibnamefont {Wambach}},\ }in\
  \href@noop {} {\emph {\bibinfo {booktitle} {{Compact Stars in the QCD Phase
  Diagram IV}}}}\ (\bibinfo {year} {2015})\ \Eprint
  {http://arxiv.org/abs/1506.04260} {arXiv:1506.04260 [hep-ph]} \BibitemShut
  {NoStop}%
\bibitem [{\citenamefont {Lee}\ \emph {et~al.}(2015)\citenamefont {Lee},
  \citenamefont {Nakano}, \citenamefont {Tsue}, \citenamefont {Tatsumi},\ and\
  \citenamefont {Friman}}]{Lee:2015bva}%
  \BibitemOpen
  \bibfield  {author} {\bibinfo {author} {\bibfnamefont {T.-G.}\ \bibnamefont
  {Lee}}, \bibinfo {author} {\bibfnamefont {E.}~\bibnamefont {Nakano}},
  \bibinfo {author} {\bibfnamefont {Y.}~\bibnamefont {Tsue}}, \bibinfo {author}
  {\bibfnamefont {T.}~\bibnamefont {Tatsumi}}, \ and\ \bibinfo {author}
  {\bibfnamefont {B.}~\bibnamefont {Friman}},\ }\href {\doibase
  10.1103/PhysRevD.92.034024} {\bibfield  {journal} {\bibinfo  {journal} {Phys.
  Rev. D}\ }\textbf {\bibinfo {volume} {92}},\ \bibinfo {pages} {034024}
  (\bibinfo {year} {2015})},\ \Eprint {http://arxiv.org/abs/1504.03185}
  {arXiv:1504.03185 [hep-ph]} \BibitemShut {NoStop}%
\bibitem [{\citenamefont {Hidaka}\ \emph {et~al.}(2015)\citenamefont {Hidaka},
  \citenamefont {Kamikado}, \citenamefont {Kanazawa},\ and\ \citenamefont
  {Noumi}}]{Hidaka:2015xza}%
  \BibitemOpen
  \bibfield  {author} {\bibinfo {author} {\bibfnamefont {Y.}~\bibnamefont
  {Hidaka}}, \bibinfo {author} {\bibfnamefont {K.}~\bibnamefont {Kamikado}},
  \bibinfo {author} {\bibfnamefont {T.}~\bibnamefont {Kanazawa}}, \ and\
  \bibinfo {author} {\bibfnamefont {T.}~\bibnamefont {Noumi}},\ }\href
  {\doibase 10.1103/PhysRevD.92.034003} {\bibfield  {journal} {\bibinfo
  {journal} {Phys. Rev. D}\ }\textbf {\bibinfo {volume} {92}},\ \bibinfo
  {pages} {034003} (\bibinfo {year} {2015})},\ \Eprint
  {http://arxiv.org/abs/1505.00848} {arXiv:1505.00848 [hep-ph]} \BibitemShut
  {NoStop}%
\bibitem [{\citenamefont {Yoshiike}\ \emph {et~al.}(2017)\citenamefont
  {Yoshiike}, \citenamefont {Lee},\ and\ \citenamefont
  {Tatsumi}}]{Yoshiike:2017kbx}%
  \BibitemOpen
  \bibfield  {author} {\bibinfo {author} {\bibfnamefont {R.}~\bibnamefont
  {Yoshiike}}, \bibinfo {author} {\bibfnamefont {T.-G.}\ \bibnamefont {Lee}}, \
  and\ \bibinfo {author} {\bibfnamefont {T.}~\bibnamefont {Tatsumi}},\ }\href
  {\doibase 10.1103/PhysRevD.95.074010} {\bibfield  {journal} {\bibinfo
  {journal} {Phys. Rev. D}\ }\textbf {\bibinfo {volume} {95}},\ \bibinfo
  {pages} {074010} (\bibinfo {year} {2017})},\ \Eprint
  {http://arxiv.org/abs/1702.01511} {arXiv:1702.01511 [hep-ph]} \BibitemShut
  {NoStop}%
\bibitem [{\citenamefont {Pisarski}\ \emph {et~al.}(2019)\citenamefont
  {Pisarski}, \citenamefont {Skokov},\ and\ \citenamefont
  {Tsvelik}}]{Pisarski:2018bct}%
  \BibitemOpen
  \bibfield  {author} {\bibinfo {author} {\bibfnamefont {R.~D.}\ \bibnamefont
  {Pisarski}}, \bibinfo {author} {\bibfnamefont {V.~V.}\ \bibnamefont
  {Skokov}}, \ and\ \bibinfo {author} {\bibfnamefont {A.~M.}\ \bibnamefont
  {Tsvelik}},\ }\href {\doibase 10.1103/PhysRevD.99.074025} {\bibfield
  {journal} {\bibinfo  {journal} {Phys. Rev. D}\ }\textbf {\bibinfo {volume}
  {99}},\ \bibinfo {pages} {074025} (\bibinfo {year} {2019})},\ \Eprint
  {http://arxiv.org/abs/1801.08156} {arXiv:1801.08156 [hep-ph]} \BibitemShut
  {NoStop}%
\bibitem [{\citenamefont {Pisarski}\ \emph {et~al.}(2020)\citenamefont
  {Pisarski}, \citenamefont {Tsvelik},\ and\ \citenamefont
  {Valgushev}}]{Pisarski:2020dnx}%
  \BibitemOpen
  \bibfield  {author} {\bibinfo {author} {\bibfnamefont {R.~D.}\ \bibnamefont
  {Pisarski}}, \bibinfo {author} {\bibfnamefont {A.~M.}\ \bibnamefont
  {Tsvelik}}, \ and\ \bibinfo {author} {\bibfnamefont {S.}~\bibnamefont
  {Valgushev}},\ }\href {\doibase 10.1103/PhysRevD.102.016015} {\bibfield
  {journal} {\bibinfo  {journal} {Phys. Rev. D}\ }\textbf {\bibinfo {volume}
  {102}},\ \bibinfo {pages} {016015} (\bibinfo {year} {2020})},\ \Eprint
  {http://arxiv.org/abs/2005.10259} {arXiv:2005.10259 [hep-ph]} \BibitemShut
  {NoStop}%
\bibitem [{\citenamefont {Tripolt}\ \emph {et~al.}(2018)\citenamefont
  {Tripolt}, \citenamefont {Schaefer}, \citenamefont {von Smekal},\ and\
  \citenamefont {Wambach}}]{Tripolt:2017zgc}%
  \BibitemOpen
  \bibfield  {author} {\bibinfo {author} {\bibfnamefont {R.-A.}\ \bibnamefont
  {Tripolt}}, \bibinfo {author} {\bibfnamefont {B.-J.}\ \bibnamefont
  {Schaefer}}, \bibinfo {author} {\bibfnamefont {L.}~\bibnamefont {von
  Smekal}}, \ and\ \bibinfo {author} {\bibfnamefont {J.}~\bibnamefont
  {Wambach}},\ }\href {\doibase 10.1103/PhysRevD.97.034022} {\bibfield
  {journal} {\bibinfo  {journal} {Phys. Rev. D}\ }\textbf {\bibinfo {volume}
  {97}},\ \bibinfo {pages} {034022} (\bibinfo {year} {2018})},\ \Eprint
  {http://arxiv.org/abs/1709.05991} {arXiv:1709.05991 [hep-ph]} \BibitemShut
  {NoStop}%
\bibitem [{\citenamefont {Wetterich}(1993)}]{Wetterich:1992yh}%
  \BibitemOpen
  \bibfield  {author} {\bibinfo {author} {\bibfnamefont {C.}~\bibnamefont
  {Wetterich}},\ }\href {\doibase 10.1016/0370-2693(93)90726-X} {\bibfield
  {journal} {\bibinfo  {journal} {Phys. Lett. B}\ }\textbf {\bibinfo {volume}
  {301}},\ \bibinfo {pages} {90} (\bibinfo {year} {1993})},\ \Eprint
  {http://arxiv.org/abs/1710.05815} {arXiv:1710.05815 [hep-th]} \BibitemShut
  {NoStop}%
\end{thebibliography}%
\bibliographystyle{apsrev4-1}

\end{document}